\documentclass[12pt]{article}
\usepackage{epsfig}
\usepackage{cite}
\usepackage{xcolor}
\usepackage{hyperref}

\def\beq{\begin{equation}}
\def\eeq{\end{equation}}
\def\beqa{\begin{eqnarray}}
\def\eeqa{\end{eqnarray}}
 
\newlength{\dinwidth} \newlength{\dinmargin}
\begin{document}

\begin{center}
{\Large \bf Top-quark cross sections and distributions at aN$^3$LO}
\end{center}
\vspace{2mm}
\begin{center}
{\large Nikolaos Kidonakis, Marco Guzzi, Alberto Tonero}\\
\vspace{2mm}
{\it Department of Physics, Kennesaw State University,\\
Kennesaw, GA 30144, USA}
\end{center}
 
\begin{abstract}
We present theoretical predictions for top-quark total cross sections and for differential distributions in top-quark transverse momentum and rapidity. Third-order soft gluon corrections are added to the complete NNLO quantities to provide approximate N$^3$LO (aN$^3$LO) results, and electroweak corrections at NLO are also included. 
We calculate the theoretical uncertainties from scale dependence and from parton distribution functions in the proton, and estimate their impact on the total and differential cross sections. The higher-order corrections are large and they reduce the scale uncertainties. The results presented in this work include the best current theoretical input and are in good agreement with recent data from the LHC.
\end{abstract}

\section{Introduction}

The study of the top quark is a central topic in current research in particle physics due to the unique properties of the top quark and its potential role in the search for new physics and in placing constraints on parton distribution functions (pdf) of the proton. Top-antitop production is the dominant mode at LHC energies. Total cross sections and differential distributions in $t{\bar t}$ production have been calculated theoretically and measured experimentally at the Tevatron and the LHC to a high precision (see Refs. \cite{NKtoprev,SNHQ} for reviews). 

Theoretical calculations for top-quark cross sections and differential distributions in $t{\bar t}$ production were performed at next-to-leading order (NLO) in Refs. \cite{NLOtt1,NLOtt2,NLOtt3,NLOtt4}, and at next-to-next-to-leading order (NNLO) in Refs. \cite{Czakon:2013goa,NNLOttLHC,NNLOttTev,Catani:2019iny,Catani:2019hip}. Electroweak corrections at NLO  were calculated in Refs. \cite{EW1,EW2,EW3,EW4,EW5,EW6}. 

At each perturbative order in QCD, the cross section receives contributions from soft-gluon emission which are particularly important near partonic threshold and which can be formally resummed. Higher-order soft-gluon corrections for $t{\bar t}$ differential cross sections were calculated from QCD threshold resummation at leading-logarithmic (LL) accuracy in Ref. \cite{NKJS}, at next-to-leading-logarithm (NLL) accuracy in Refs. \cite{NKGS1,NKGS2,NKnll}, and at next-to-next-to-leading-logarithm (NNLL) accuracy in  Refs. \cite{NK2loop,NKnnll1,NKnnll2,NKnnll3,NKan3lo1,NKan3lo2,NKan3lo3,NKdoublediff}. We use the theoretical work of Refs. \cite{NK2loop,NKnnll1,NKnnll2,NKnnll3,NKan3lo1,NKan3lo2,NKan3lo3,NKdoublediff} on soft-gluon resummation at NNLL accuracy in the results that we present here. A discussion of other resummation approaches can be found in the reviews of Refs. \cite{NKtoprev,NKBP}. 

The soft-gluon corrections are an important subset of the QCD corrections, and they are numerically dominant at LHC energies \cite{NKtoprev}. These soft corrections in $t{\bar t}$ production provide excellent approximations at NLO and NNLO to the complete set of QCD corrections \cite{NKtoprev} and, in fact, they predicted the NNLO results with high precision and accuracy for both the total cross section and the top-quark differential distributions in transverse momentum and rapidity \cite{NKnnll1,NKnnll2,NKnnll3}. In going beyond the NNLO results, the soft-gluon corrections provide significant enhancements and a reduction of the scale dependence when calculated at next-to-next-to-next-to-leading order (N$^3$LO) \cite{NKan3lo1,NKan3lo2,NKan3lo3,NKdoublediff}. Fixed-order expansions of resummed cross sections do not require prescriptions and are, therefore, preferred for the reasons detailed in Refs. \cite{NKtoprev,NKnll} and because they have been very successful in approximating the complete corrections at NLO and NNLO, as noted above.

In this paper, we use the formalism of Refs. \cite{NKGS1,NKGS2,NKnll,NK2loop,NKnnll1,NKnnll2,NKnnll3,NKan3lo1,NKan3lo2,NKan3lo3,NKdoublediff}, and in particular Refs. \cite{NKan3lo1,NKan3lo2}, to calculate soft-gluon corrections at third order in perturbative QCD, and we add them to the exact NNLO results. We denote these results with exact NNLO plus third-order soft-gluon corrections as approximate N$^3$LO (aN$^3$LO). Furthermore, we include electroweak (EW) corrections at NLO. We provide updated and new results for total $t{\bar t}$ cross sections at LHC energies, and also new results for binned top-quark differential distributions in transverse momentum and rapidity.

In Sec.~\ref{Soft-gluon corrections and resummation}, we give a brief overview of the resummation formalism for $t{\bar t}$ production. In Sec.~\ref{Total ttbar cross sections}, we provide results for the total cross section at LHC energies. Sec.~\ref{pTt distributions at 13 TeV} has results for the top-quark transverse-momentum distributions and Sec.~\ref{yt distributions at 13 TeV} has the corresponding results for the top-quark rapidity distributions. We conclude in Sec.~\ref{Conclusion}.

\section{Soft-gluon corrections and resummation}
\label{Soft-gluon corrections and resummation}

In this section, we give a brief overview of the resummation formalism for the calculation of soft-gluon corrections in $t{\bar t}$ production \cite{NKGS1,NKGS2,NKnll,NK2loop,NKnnll1,NKnnll2,NKnnll3,NKan3lo1,NKan3lo2,NKan3lo3,NKdoublediff} (see \cite{NKresrev} for a review). The origin of these corrections is from the emission of soft, i.e. low-energy, gluons, resulting in partial cancellations of infrared divergences between real-emission and virtual diagrams near partonic threshold.

We consider partonic processes, 
\beq
f_{1}(p_1)\, + \, f_{2}\, (p_2) \rightarrow t(p_t) \, + \, {\bar t}(p_{\bar t})  \, ,
\eeq
where $f_1$ and $f_2$ represent quarks and/or gluons in the colliding protons. At leading order (LO), the partonic channels are $q{\bar q} \to t{\bar t}$ and $gg \to t{\bar t}$. We define the standard partonic kinematical variables from the 4-momenta of the particles, $s=(p_1+p_2)^2$, $t=(p_1-p_t)^2$, and $u=(p_2-p_t)^2$. We define a partonic threshold variable $s_4=s+t+u-2m_t^2$, with $m_t$ the top-quark mass, or equivalently $s_4=(p_{\bar t}+p_g)^2-m_t^2$, where $p_g$ is the momentum of an additional gluon in the final state. As $p_g \to 0$, i.e. as we approach partonic threshold, we have $s_4 \to 0$. We remark that partonic threshold is a generalized definition of threshold: the $t{\bar t}$ pair is not necessarily produced at rest.
 
The resummation, i.e. exponentiation, of soft-gluon corrections follows from factorization properties of the double-differential cross section under Laplace transforms. The partonic cross section can be expressed as a product of different functions which describe the emission of soft and collinear quanta as well as the hard scattering \cite{NKGS1,NKGS2,NKnll,NK2loop,NKnnll1,LOS}. The renormalization-group evolution of the soft function, which describes noncollinear soft-gluon emission, involves a soft anomalous dimension matrix for each partonic channel, and it results in the exponentiation of the transform variable conjugate to $s_4$.  

The formalism for the derivation of soft-gluon resummation from factorization follows the steps below. We first write the factorized form for the differential hadronic cross section,  $d\sigma_{pp \to t{\bar t}}$, as a convolution of the differential partonic cross section, $d{\hat \sigma}_{ab \to t{\bar t}}$, with the pdf $\phi_{a/p}$ and $\phi_{b/p}$, as 
\beq
d\sigma_{pp \to t{\bar t}}=\sum_{a,b} \; 
\int dx_a \, dx_b \,  \phi_{a/p}(x_a, \mu_F) \, \phi_{b/p}(x_b, \mu_F)  \, 
d{\hat \sigma}_{ab \to t{\bar t}}(s_4, \mu_F)  \, ,
\label{fact1}
\eeq
where $\mu_F$ is the factorization scale, and $x_a$, $x_b$ are momentum fractions of partons $a$, $b$, respectively,  in the colliding protons.

We take Laplace transforms, with transform variable $N$, of the partonic cross section via $d{\tilde{\hat\sigma}}_{ab \to t{\bar t}}(N)=\int_0^s (ds_4/s) \,  e^{-N s_4/s} \, d{\hat\sigma}_{ab \to t{\bar t}}(s_4)$, and of the pdf as ${\tilde \phi}(N)=\int_0^1 e^{-N(1-x)} \phi(x) \, dx$. Then, under transforms, Eq. (\ref{fact1}) gives the $N$-space expression at the parton level
\beq
d{\tilde \sigma}_{ab \to t{\bar t}}(N)= {\tilde \phi}_{a/a}(N_a, \mu_F) \, {\tilde \phi}_{b/b}(N_b, \mu_F) \, d{\tilde{\hat \sigma}}_{ab \to t{\bar t}}(N, \mu_F) \, .
\label{fact1N}
\eeq

A further refactorization of the cross section is possible and is written in terms of a short-distance hard function $H_{ab \to t{\bar t}}$, a soft function $S_{ab \to t{\bar t}}$ for noncollinear soft-gluon emission, and distributions $\psi_{i/i}$ for collinear gluon emission from the incoming partons:
\beq
d{\tilde{\sigma}}_{ab \to t{\bar t}}(N)={\tilde \psi}_{a/a}(N_a,\mu_F) \, {\tilde \psi}_{b/b}(N_b,\mu_F) \, {\rm tr} \left\{H_{ab \to t{\bar t}} \left(\alpha_s(\mu_R)\right) \, {\tilde S}_{ab \to t{\bar t}} \left(\frac{\sqrt{s}}{N \mu_F} \right)\right\} \, .
\label{fact2N}
\eeq
We note that  $H_{ab \to t{\bar t}}$ and $S_{ab \to t{\bar t}}$ are process-dependent matrices in the space of color exchanges in the hard scattering. They are $2 \times 2$ matrices for $q{\bar q} \to t{\bar t}$ , and $3 \times 3$ matrices for $gg \to t{\bar t}$, and we take the trace of their product.

Using Eqs. (\ref{fact1N}) and (\ref{fact2N}), we get an expression for the partonic cross section 
\beq
d{\tilde{\hat \sigma}}_{ab \to t{\bar t}}(N,\mu_F)=
\frac{{\tilde \psi}_{a/a}(N_a, \mu_F) \, {\tilde \psi}_{b/b}(N_b, \mu_F)}{{\tilde \phi}_{a/a}(N_a, \mu_F) \, {\tilde \phi}_{b/b}(N_b, \mu_F)} \; \,  {\rm tr} \left\{H_{ab \to t{\bar t}}\left(\alpha_s(\mu_R)\right) \, {\tilde S}_{ab \to t{\bar t}}\left(\frac{\sqrt{s}}{N \mu_F} \right)\right\} \, .
\label{partonN}
\eeq

The renormalization-group evolution of the $N$-dependent functions in Eq. (\ref{partonN}) leads to resummation, i.e. exponentiation, of the corrections from collinear and soft gluons. The resummed cross section is given by 
\beqa
d{\tilde{\hat \sigma}}_{ab \to t{\bar t}}^{\rm resum}(N,\mu_F) &=&
\exp\left[\sum_{i=a,b} E_{i}(N_i)\right] \, 
\exp\left[\sum_{i=a,b} 2 \int_{\mu_F}^{\sqrt{s}} \frac{d\mu}{\mu} \gamma_{i/i}(N_i)\right] 
\nonumber \\ && \hspace{-2mm} 
\times {\rm tr} \left\{H_{ab \to t{\bar t}}\left(\alpha_s(\sqrt{s})\right) {\bar P} \, \exp \left[\int_{\sqrt{s}}^{{\sqrt{s}}/N} \frac{d\mu}{\mu} \, \Gamma_{\! S \, ab \to t{\bar t}}^{\dagger} \left(\alpha_s(\mu)\right)\right] \right.
\nonumber \\ && \hspace{8mm}
\left. 
\times {\tilde S}_{ab \to t{\bar t}} \left(\alpha_s\left(\frac{\sqrt{s}}{N}\right)\right) \;
P\, \exp \left[\int_{\sqrt{s}}^{{\sqrt{s}}/N} \frac{d\mu}{\mu} \, \Gamma_{\! S \, ab \to t{\bar t}} \left(\alpha_s(\mu)\right)\right] \right\} \, .
\label{resummed}
\eeqa
Here, $P$ (${\bar P}$) denotes path ordering in the same (reverse) sense as the integration variable $\mu$. The first exponential in Eq. (\ref{resummed}) resums collinear and soft contributions from the incoming partons, and it involves universal functions that depend only on whether those partons are quarks or gluons \cite{GS87,CT89}. The second exponential expresses the factorization-scale dependence in terms of the anomalous dimension $\gamma_{i/i}$ of the parton density $\phi_{i/i}$. The resummation of noncollinear soft-gluon emission is performed via the soft anomalous dimensions $\Gamma_{\! S \, q{\bar q} \to t{\bar t}}$, which are $2 \times 2$ matrices, and $\Gamma_{\! S \, gg \to t{\bar t}}$, which are $3 \times 3$ matrices. The matrices $\Gamma_{\! S \, q{\bar q} \to t{\bar t}}$ and $\Gamma_{\! S \, gg \to t{\bar t}}$ are known at one loop \cite{NKGS1,NKGS2} and two loops \cite{NK2loop,FNPY,NKnnll1}. Partial results for $\Gamma_S$ also exist at three loops \cite{NKresrev}, and the recent calculation \cite{NK4loop} of the four-loop massive cusp anomalous dimension from its asymptotics also provides partial contributions to the four-loop $\Gamma_S$.

After doing the inverse transform at fixed perturbative order, the soft-gluon corrections take the form of plus distributions of logarithms of $s_4$. Specifically, the soft-gluon contributions involve terms of the form $[(\ln^k(s_4/m_t^2))/s_4]_+$, with $0 \le k \le 2n-1$ at $n$th order in the strong coupling, $\alpha_s$. We use the resummed cross section as a generator of fixed-order results via expansions which do not require prescriptions. We prefer this method for the reasons detailed in Refs. \cite{NKtoprev,NKnll}, specifically to avoid underestimates of the size of the corrections, and because such expansions have been consistently very successful in approximating and predicting the complete corrections at NLO and NNLO, as has long been demonstrated \cite{NKtoprev,NKnnll1,NKnnll2,NKnnll3}.

We refer the reader to Refs. \cite{NKtoprev,NKGS1,NKGS2,NKnll,NK2loop,NKnnll1,NKnnll2,NKnnll3,NKan3lo1,NKan3lo2,NKan3lo3,NKdoublediff,NKresrev} for more details on the formalism and for past applications to $t{\bar t}$ production.

\section{Total $t{\bar t}$ cross sections}
\label{Total ttbar cross sections}

In this section we present theoretical predictions for the total cross section of $t \bar t$ production at the LHC, for different values of the collider energy and up to aN$^3$LO in QCD with the inclusion of NLO EW corrections. We set the factorization and renormalization scales equal to a common scale denoted by $\mu$, namely $\mu_F=\mu_R=\mu$. The central results are obtained by setting $\mu=m_t$, where the top-quark mass in the pole-mass approximation is taken to be $m_t$ = 172.5 GeV. Scale uncertainties are obtained by varying the common scale $\mu$ in the range $m_t/2\leq \mu\leq 2 m_t$. We checked that uncertainties obtained from the envelope of a 7-point scale variation, where $\mu_F$ and $\mu_R$ are varied independently, are basically the same as the ones obtained by performing a simpler 3-point scale variation.

The results at LO and NLO QCD are calculated using in-house codes, and they are also cross-checked with \texttt{Top++2.0} \cite{Czakon:2011xx} and \texttt{MadGraph5 aMC@NLO} \cite{MG5}. The results at NNLO QCD are calculated using \texttt{Top++2.0}, and we find that they are very close (at the per mille level) to those at approximate NNLO (aNNLO) QCD, where aNNLO denotes the sum of NLO and the second-order soft-gluon corrections. This, again, shows that the higher-order QCD corrections are dominated numerically by soft-gluon contributions, as has been known for a long time for $t{\bar t}$ production \cite{NKtoprev,NKnll,NKnnll1,NKnnll2,NKnnll3,NKan3lo1,NKan3lo2,NKan3lo3,NKdoublediff}. 

To determine the aN$^3$LO QCD corrections (i.e. the third-order soft-gluon corrections) we use the analytical results of Ref. \cite{NKan3lo1}. These aN$^3$LO QCD corrections are then added to the NNLO QCD result to derive the aN$^3$LO QCD total cross section.

In addition, we use \texttt{MadGraph5 aMC@NLO} \cite{MG5,MGew} to compute NLO QCD+EW cross sections. From those we determine the magnitude of NLO EW corrections which are added to the NNLO and aN$^3$LO QCD cross sections to obtain NNLO QCD+EW and aN$^3$LO QCD+EW results, respectively. The difference between the NLO QCD and NLO QCD+EW cross sections is of the order of 0.1\% or less at 5.02 TeV and grows to 0.4\% at 13, 13.6 and 14 TeV, making the EW effects more visible at higher collision energies. 

We perform calculations using four different pdf sets: MSHT20 NNLO \cite{MSHT20}, MSHT20 aN$^3$LO \cite{MSHT20a}, CT18 NNLO \cite{CT18}, and NNPDF4.0 NNLO \cite{NNPDF}. We show results for each perturbative QCD order from LO through aN$^3$LO, including EW corrections, in Tables \ref{table1}-\ref{table4}. We use the same pdf set for computing results at every perturbative order, in order to show how each order in the series contributes to the aN$^3$LO total cross section. In addition, for each pdf set we provide pdf uncertainties. We find that scale uncertainties decrease substantially with the increase of the perturbative order. Moreover, looking at each order, one sees that both scale and pdf uncertainties slightly decrease with the increase of the collider energy.

\begin{table}[htb]
\begin{center}
\begin{tabular}{|c|c|c|c|c|c|c|} \hline
\multicolumn{7}{|c|}{$t{\bar t}$ total cross sections at LHC energies with MSHT20 NNLO pdf} \\ \hline
$\sigma$ in pb & 5.02 TeV & 7 TeV & 8 TeV & 13 TeV & 13.6 TeV & 14 TeV \\ \hline
LO QCD & $40.9^{+15.5}_{-10.4}{}^{+1.2}_{-0.8}$ & $105^{+37}_{-25}{}^{+3}_{-2}$ & $150^{+50}_{-35}{}^{+4}_{-2}$ & $487^{+142}_{-103}{}^{+10}_{-6}$ & $540^{+155}_{-113}{}^{+10}_{-7}$ & $576^{+163}_{-120}{}^{+11}_{-7}$ \\ \hline
NLO QCD &$59.6^{+7.1}_{-8.1}{}^{+2.0}_{-1.2}$& $155^{+19}_{-20}{}^{+4}_{-3}$ &$222^{+26}_{-28}{}^{+6}_{-4}$ & $730^{+85}_{-86}{}^{+14}_{-10}$ & $809^{+94}_{-94}{}^{+16}_{-11}$ &$863^{+101}_{-99}{}^{+17}_{-11}$ \\ \hline
NLO QCD+EW & $59.6^{+7.0}_{-8.1}{}^{+1.9}_{-1.2}$ & $155^{+18}_{-20}{}^{+4}_{-3}$ & $221^{+26}_{-28}{}^{+6}_{-3}$ & $727^{+83}_{-85}{}^{+14}_{-10}$ & $806^{+92}_{-93}{}^{+15}_{-11}$ & $860^{+99}_{-99}{}^{+17}_{-11}$ \\ \hline
NNLO QCD& $67.1^{+3.0}_{-4.6}{}^{+2.2}_{-1.4}$ & $174^{+7}_{-11}{}^{+5}_{-3}$ & $249^{+10}_{-16}{}^{+7}_{-4}$ & $814^{+28}_{-46}{}^{+16}_{-11}$ & $902^{+31}_{-50}{}^{+18}_{-12}$ & $963^{+33}_{-53}{}^{+18}_{-13}$ \\ \hline
NNLO QCD+EW & $67.1^{+3.0}_{-4.6}{}^{+2.2}_{-1.4}$ & $174^{+7}_{-11}{}^{+5}_{-3}$ & $248^{+10}_{-16}{}^{+7}_{-4}$ & $811^{+28}_{-46}{}^{+16}_{-11}$ & $899^{+31}_{-50}{}^{+18}_{-12}$ & $960^{+33}_{-53}{}^{+18}_{-13}$ \\ \hline
aN$^3$LO QCD& $70.2^{+2.2}_{-3.3}{}^{+2.3}_{-1.5}$ & $181^{+5}_{-7}{}^{+5}_{-3}$ & $258^{+7}_{-9}{}^{+7}_{-4}$ & $839^{+23}_{-18}{}^{+17}_{-11}$ & $928^{+25}_{-20}{}^{+18}_{-12}$ & $990^{+27}_{-22}{}^{+19}_{-13}$ \\ \hline
aN$^3$LO QCD+EW & $70.2^{+2.2}_{-3.3}{}^{+2.3}_{-1.5}$ & $181^{+5}_{-7}{}^{+5}_{-3}$ & $257^{+7}_{-9}{}^{+7}_{-4}$ & $836^{+23}_{-18}{}^{+17}_{-11}$ & $925^{+25}_{-20}{}^{+18}_{-12}$ & $987^{+27}_{-22}{}^{+19}_{-13}$ \\ \hline
\end{tabular}
\caption[]{The $t{\bar t}$ total cross sections (in pb, with central result for $\mu=m_t$, and uncertainties from scale variation and pdf) at different perturbative orders in $pp$ collisions at the LHC with various values of $\sqrt{S}$, with $m_t=172.5$ GeV and MSHT20 NNLO pdf.}
\label{table1}
\end{center}
\end{table}

In Table~\ref{table1} we show the total rates for $t {\bar t}$ production, together with scale and pdf uncertainties, for various LHC energies at LO, NLO, NNLO and aN$^3$LO QCD, and also with NLO EW corrections, using MSHT20 NNLO pdf (see also \cite{NKsnowmass} for aN$^3$LO QCD results). At NNLO QCD(+EW), the scale uncertainty varies from $+4.5$\% $-6.9$\% at 5.02 TeV to $+3.4$\% $-5.5$\% at 14 TeV. At aN$^3$LO QCD(+EW), the scale uncertainty varies from $+3.1$\% $-4.7$\% at 5.02 TeV to $+2.7$ $-2.2$\% at 14 TeV. At both NNLO and aN$^3$LO QCD(+EW), the pdf uncertainty varies from $+3.3$\% $-2.1$\% at 5.02 TeV to $+1.9$\% $-1.3$\% at 14 TeV.

The QCD $K$-factors are large, showing the importance of the higher-order QCD corrections. For example, at 13 TeV the NLO/LO $K$-factor is 1.50, the NNLO/LO $K$-factor is 1.67, and the aN$^3$LO/LO $K$-factor is 1.72. Including EW corrections does not materially modify the $K$-factors.

\begin{table}[htb]
\begin{center}
\begin{tabular}{|c|c|c|c|c|c|c|} \hline
\multicolumn{7}{|c|}{$t{\bar t}$ total cross sections at LHC energies with MSHT20 aN$^3$LO pdf} \\ \hline
$\sigma$ in pb & 5.02 TeV & 7 TeV & 8 TeV & 13 TeV & 13.6 TeV & 14 TeV \\ \hline
LO  QCD& $40.0^{+14.9}_{-10.1}{}^{+1.1}_{-1.2}$ & $103^{+35}_{-24}{}^{+3}_{-3}$ & $146^{+48}_{-34}{}^{+3}_{-4}$ & $469^{+133}_{-97}{}^{+9}_{-10}$ & $518^{+145}_{-106}{}^{+10}_{-11}$ & $553^{+153}_{-113}{}^{+11}_{-11}$ \\ \hline
NLO  QCD&$58.1^{+6.8}_{-7.8}{}^{+1.8}_{-2.0}$ & $151^{+17}_{-20}{}^{+4}_{-5}$ & $215^{+25}_{-27}{}^{+5}_{-6}$ &$700^{+80}_{-80}{}^{+15}_{-15}$ & $775^{+89}_{-88}{}^{+16}_{-16}$ &$828^{+94}_{-94}{}^{+16}_{-18}$ \\ \hline
NLO  QCD+EW & $58.1^{+6.6}_{-7.8}{}^{+1.8}_{-2.0}$ & $150^{+17}_{-19}{}^{+4}_{-4}$ & $214^{+25}_{-26}{}^{+6}_{-6}$ & $698^{+78}_{-80}{}^{+14}_{-16}$ & $772^{+88}_{-87}{}^{+16}_{-16}$ & $825^{+92}_{-93}{}^{+16}_{-18}$ \\ \hline
NNLO  QCD& $65.3^{+2.8}_{-4.4}{}^{+2.0}_{-2.2}$ & $169^{+7}_{-11}{}^{+5}_{-5}$ & $240^{+9}_{-15}{}^{+6}_{-7}$ & $781^{+27}_{-43}{}^{+16}_{-17}$ & $864^{+30}_{-47}{}^{+18}_{-19}$ & $922^{+32}_{-49}{}^{+18}_{-20}$ \\ \hline
NNLO  QCD+EW & $65.3^{+2.8}_{-4.4}{}^{+2.0}_{-2.2}$ & $168^{+7}_{-11}{}^{+5}_{-5}$ & $239^{+9}_{-15}{}^{+6}_{-7}$ & $779^{+27}_{-43}{}^{+16}_{-17}$ & $861^{+30}_{-47}{}^{+18}_{-19}$ & $919^{+32}_{-49}{}^{+18}_{-20}$ \\ \hline
aN$^3$LO  QCD& $68.2^{+2.1}_{-3.2}{}^{+2.1}_{-2.3}$ & $175^{+5}_{-7}{}^{+5}_{-5}$ & $249^{+7}_{-9}{}^{+6}_{-7}$ & $804^{+22}_{-17}{}^{+16}_{-17}$ & $889^{+24}_{-19}{}^{+18}_{-20}$ & $948^{+26}_{-21}{}^{+19}_{-21}$ \\ \hline
aN$^3$LO  QCD+EW & $68.2^{+2.1}_{-3.2}{}^{+2.1}_{-2.3}$ & $174^{+5}_{-7}{}^{+5}_{-5}$ & $248^{+7}_{-9}{}^{+6}_{-7}$ & $802^{+22}_{-17}{}^{+16}_{-17}$ & $886^{+24}_{-19}{}^{+18}_{-20}$ & $945^{+26}_{-21}{}^{+19}_{-21}$ \\ \hline
\end{tabular}
\caption[]{The $t{\bar t}$ total cross sections (in pb, with central result for $\mu=m_t$, and uncertainties from scale variation and pdf) at different perturbative orders in $pp$ collisions at the LHC with various values of $\sqrt{S}$, with $m_t=172.5$ GeV and MSHT20 aN$^3$LO pdf.}
\label{table2}
\end{center}
\end{table}

In Table~\ref{table2} we show the corresponding total rates for $t \bar t$ production, together with scale and pdf uncertainties, for various LHC energies using MSHT20 aN$^3$LO pdf. At both NNLO QCD(+EW) and aN$^3$LO QCD(+EW), the relative magnitude of scale uncertainties is similar to the case of MSHT20 NNLO pdf, while pdf uncertainties are slightly bigger for MSHT20 aN$^3$LO pdf, being $+3.1$\% $-3.4$\% at 5.02 TeV and $+2.0$ $-2.2$\% at 14 TeV. The $K$-factors are also very similar to those for MSHT20 NNLO pdf.
 
\begin{table}[htb]
\begin{center}
\begin{tabular}{|c|c|c|c|c|c|c|} \hline
\multicolumn{7}{|c|}{$t{\bar t}$ total cross sections at LHC energies with CT18 NNLO pdf} \\ \hline
$\sigma$ in pb & 5.02 TeV & 7 TeV & 8 TeV & 13 TeV & 13.6 TeV & 14 TeV \\ \hline
LO  QCD& $41.2^{+15.7}_{-10.5}{}^{+2.1}_{-1.3}$ & $106^{+37}_{-25}{}^{+4}_{-2}$ & $151^{+51}_{-35}{}^{+5}_{-3}$ & $491^{+142}_{-104}{}^{+10}_{-10}$ & $543^{+156}_{-113}{}^{+11}_{-10}$ & $579^{+165}_{-120}{}^{+12}_{-11}$ \\ \hline
NLO QCD & $60.3^{+7.2}_{-8.3}{}^{+3.1}_{-2.0}$ & $157^{+19}_{-20}{}^{+6}_{-4}$ & $224^{+27}_{-28}{}^{+8}_{-5}$ & $735^{+85}_{-86}{}^{+16}_{-15}$ & $814^{+95}_{-95}{}^{+17}_{-16}$ & $869^{+101}_{-101}{}^{+18}_{-17}$ \\ \hline
NLO  QCD+EW& $60.2^{+7.2}_{-8.2}{}^{+3.1}_{-1.9}$ & $157^{+18}_{-21}{}^{+5}_{-4}$ & $224^{+26}_{-29}{}^{+7}_{-6}$ & $732^{+84}_{-85}{}^{+16}_{-14}$ & $811^{+93}_{-94}{}^{+17}_{-16}$ & $866^{+99}_{-100}{}^{+17}_{-17}$ \\ \hline
NNLO  QCD& $67.9^{+3.0}_{-4.7}{}^{+3.5}_{-2.2}$ & $176^{+7}_{-11}{}^{+7}_{-4}$ & $251^{+10}_{-15}{}^{+9}_{-6}$ & $820^{+28}_{-46}{}^{+17}_{-16}$ & $908^{+31}_{-50}{}^{+19}_{-18}$ & $969^{+33}_{-53}{}^{+19}_{-18}$ \\ \hline
NNLO  QCD+EW & $67.8^{+3.0}_{-4.7}{}^{+3.5}_{-2.2}$ & $176^{+7}_{-11}{}^{+7}_{-4}$ & $251^{+10}_{-15}{}^{+9}_{-6}$ & $817^{+28}_{-46}{}^{+17}_{-16}$ & $905^{+31}_{-50}{}^{+19}_{-18}$ & $966^{+33}_{-53}{}^{{+19}}_{-18}$ \\ \hline
aN$^3$LO  QCD& $71.0^{+2.2}_{-3.3}{}^{+3.7}_{-2.3}$ & $183^{+5}_{-7}{}^{+7}_{-5}$ & $261^{+7}_{-9}{}^{+8}_{-6}$ & $845^{+23}_{-18}{}^{+18}_{-16}$ & $935^{+25}_{-20}{}^{+20}_{-18}$ & $997^{+27}_{-22}{}^{+21}_{-19}$ \\ \hline
aN$^3$LO  QCD+EW & $70.9^{+2.2}_{-3.3}{}^{+3.7}_{-2.3}$ & $183^{+5}_{-7}{}^{+7}_{-5}$ & $261^{+7}_{-9}{}^{+8}_{-6}$ & $842^{+23}_{-18}{}^{+18}_{-16}$ & $932^{+25}_{-20}{}^{+20}_{-18}$ & $994^{+27}_{-22}{}^{+21}_{-19}$ \\ \hline
\end{tabular}
\caption[]{The $t{\bar t}$ total cross sections (in pb, with central result for $\mu=m_t$, and uncertainties from scale variation and pdf) at different perturbative orders in $pp$ collisions at the LHC with various values of $\sqrt{S}$, with $m_t=172.5$ GeV and CT18 NNLO pdf.}
\label{table3}
\end{center}
\end{table}

In Table~\ref{table3} we show the total rates for $t \bar t$ production at LHC energies, together with scale and pdf uncertainties, at various perturbative orders using CT18 NNLO pdf. At both NNLO and aN$^3$LO, the relative magnitude of scale uncertainties is similar to the cases of MSHT20 NNLO and aN$^3$LO pdf. In this case, the pdf uncertainties obtained with \texttt{Top++2.0} have been rescaled by a factor of 1.645 in order to get the 68\% confidence level (C.L.) variation range for this pdf set (we note that 68\% C.L. is used for the three other pdf sets by default).
These rescaled uncertainties are somewhat bigger than MSHT20 NNLO pdf: they are $+5.2$\% $-3.2$\% at 5.02 TeV and $+ 2.0$\% $-1.9$\% at 14 TeV. The $K$-factors are very similar to those for the results in the previous two tables.

\begin{table}[htb]
\begin{center}
\begin{tabular}{|c|c|c|c|c|c|c|} \hline
\multicolumn{7}{|c|}{$t{\bar t}$ total cross sections at LHC energies with NNPDF4.0 NNLO pdf} \\ \hline
$\sigma$ in pb & 5.02 TeV & 7 TeV & 8 TeV & 13 TeV & 13.6 TeV & 14 TeV \\ \hline
LO  QCD& $38.6^{+14.5}_{-9.8}{}^{+0.4}_{-0.4}$ & $101^{+35}_{-24}{}^{+1}_{-1}$ & $144^{+49}_{-34}{}^{+1}_{-1}$ & $476^{+139}_{-101}{}^{+2}_{-3}$ & $527^{+153}_{-110}{}^{+3}_{-2}$ & $563^{+162}_{-117}{}^{+3}_{-3}$ \\ \hline
NLO  QCD & $56.1^{+6.6}_{-7.6}{}^{+0.5}_{-0.6}$ & $148^{+18}_{-19}{}^{+1}_{-1}$ & $213^{+25}_{-27}{}^{+1}_{-2}$ & $712^{+82}_{-84}{}^{+3}_{-4}$ & $790^{+91}_{-93}{}^{+4}_{-4}$ & $844^{+98}_{-98}{}^{+4}_{-5}$ \\ \hline
NLO  QCD+EW & $56.1^{+6.5}_{-7.6}{}^{+0.5}_{-0.6}$ & $148^{+17}_{-19}{}^{+1}_{-1}$ & $212^{+25}_{-26}{}^{+2}_{-1}$ & $709^{+81}_{-83}{}^{+4}_{-4}$ & $787^{+90}_{-91}{}^{+4}_{-4}$ & $841^{+96}_{-97}{}^{+4}_{-5}$ \\ \hline
NNLO  QCD& $63.1^{+2.8}_{-4.3}{}^{+0.7}_{-0.6}$ & $166^{+7}_{-10}{}^{{+2}}_{-1}$ & $239^{+9}_{-15}{}^{+2}_{-2}$ & $794^{+28}_{-44}{}^{+5}_{-4}$ & $881^{+31}_{-49}{}^{+5}_{-4}$ & $941^{+33}_{-52}{}^{+5}_{-4}$ \\ \hline
NNLO  QCD+EW & $63.1^{+2.8}_{-4.3}{}^{+0.7}_{-0.6}$ & $166^{+7}_{-10}{}^{{+2}}_{-1}$ & $238^{+9}_{-15}{}^{+2}_{-2}$ & $791^{+28}_{-44}{}^{+5}_{-4}$ & $878^{+31}_{-49}{}^{+5}_{-4}$ & $938^{+33}_{-52}{}^{+5}_{-4}$ \\ \hline
aN$^3$LO  QCD& $66.0^{+2.0}_{-3.1}{}^{+0.7}_{-0.7}$ & $173^{+5}_{-7}{}^{+2}_{-1}$ & $248^{+7}_{-9}{}^{+2}_{-2}$ & $819^{+23}_{-18}{}^{+5}_{-4}$ & $907^{+25}_{-20}{}^{+5}_{-5}$ & $969^{+27}_{-22}{}^{+5}_{-5}$ \\ \hline
aN$^3$LO  QCD+EW & $66.0^{+2.0}_{-3.1}{}^{+0.7}_{-0.7}$ & $173^{+5}_{-7}{}^{+2}_{-1}$ & $247^{+7}_{-9}{}^{+2}_{-2}$ & $816^{+23}_{-18}{}^{+5}_{-4}$ & $904^{+25}_{-20}{}^{+5}_{-5}$ & $966^{+27}_{-22}{}^{+5}_{-5}$ \\ \hline
\end{tabular}
\caption[]{The $t{\bar t}$ total cross sections (in pb, with central result for $\mu=m_t$, and uncertainties from scale variation and pdf) at different perturbative orders in $pp$ collisions at the LHC with various values of $\sqrt{S}$, with $m_t=172.5$ GeV and NNPDF4.0 NNLO pdf.}
\label{table4}
\end{center}
\end{table}

In Table~\ref{table4} we show the total rates with scale and pdf uncertainties for $t \bar t$ production for various LHC energies through aN$^3$LO QCD+EW using NNPDF4.0 NNLO pdf. At both NNLO and aN$^3$LO, the relative magnitude of scale uncertainties is similar to the previous three cases, while pdf uncertainties are much smaller for NNPDF4.0 NNLO pdf: $\pm 1.0$\% at 5.02 TeV and $\pm 0.5$\% at 14 TeV. The $K$-factors are also very similar to those for the previous three tables.

\begin{figure}[htbp]
\begin{center}
\includegraphics[width=155mm]{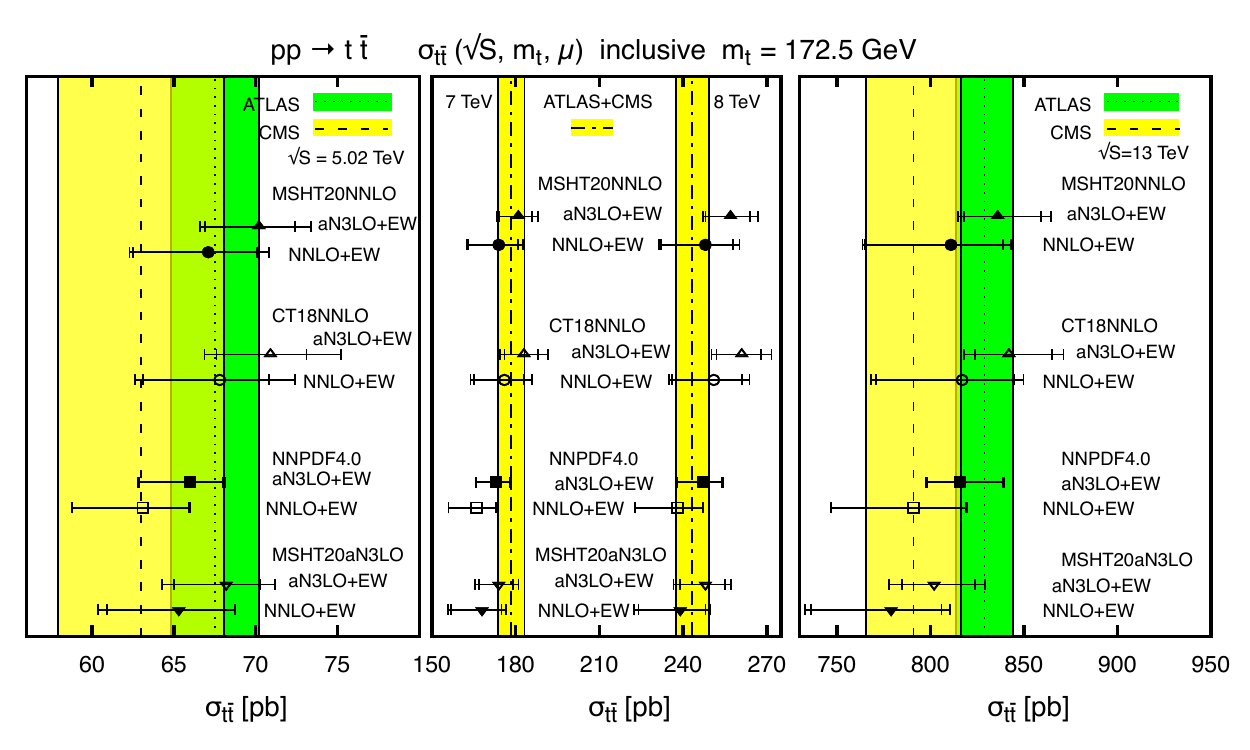}
\end{center}
\begin{center}
\includegraphics[width=155mm]{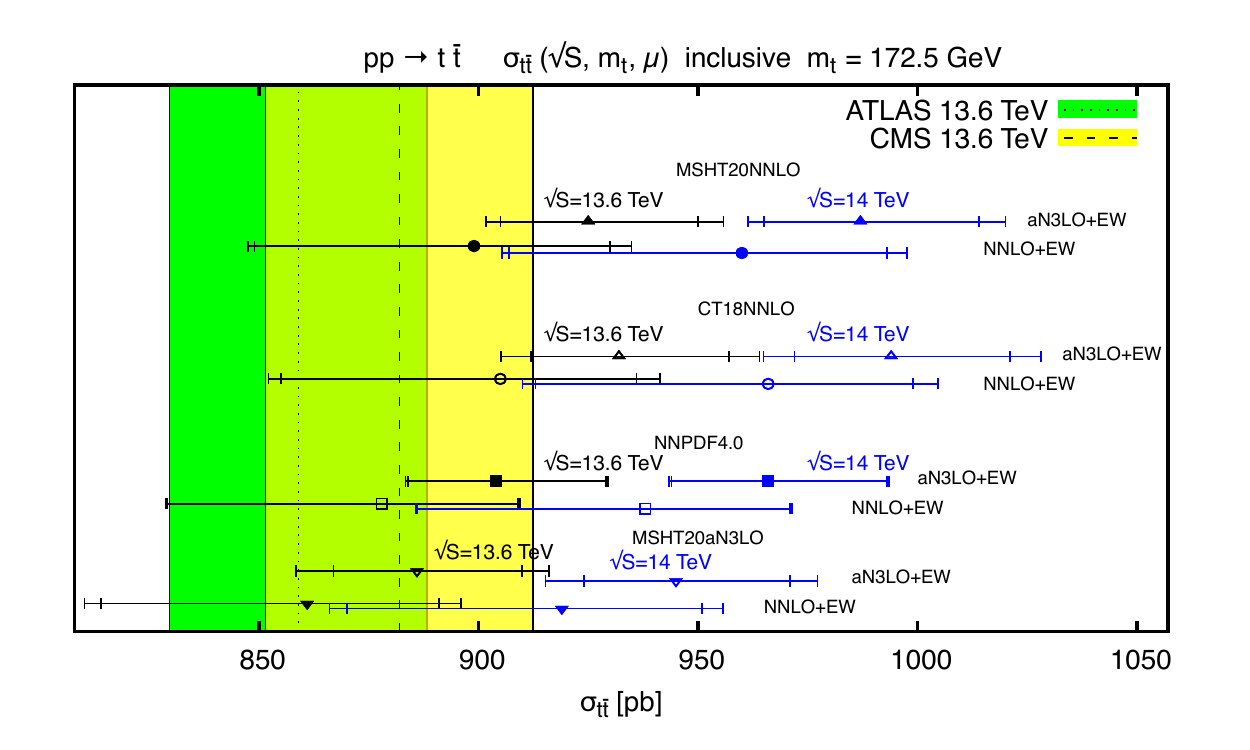}
\caption{$t \bar t$ total inclusive cross sections compared to recent measurements at the LHC at different collision energies. 
Theory error bars represent scale uncertainty (inner bar), and scale + pdf uncertainties in quadrature (outer bar). 
Experimental error bands represent all given errors added in quadrature. 
``ATLAS + CMS'' in the central inset of the upper plot indicates the most recent combination 
of ATLAS and CMS measurements at collision energies of $\sqrt{S}=$7 and 8 TeV.}  
\label{xsections}
\end{center}
\end{figure}

Finally, in Fig.~\ref{xsections} we compare our theory predictions with the most recent and accurate measurements of top-quark pair production total inclusive cross sections at the LHC. The results for the cross sections at NNLO QCD+EW and aN$^3$LO QCD+EW are shown using different pdf sets and at different collision energies. 
The smaller error bar in each theory point represents the scale uncertainty, while the larger error bar represents scale and pdf uncertainties added in quadrature. 

In the upper plot of Fig.~\ref{xsections}, the ATLAS~\cite{ATLAS:2022jbj} result of $67.5 \pm 2.7$ pb and the CMS~\cite{CMS:2021gwv} result of $63.0 \pm 5.1$ pb at collision energy $\sqrt{S}=5.02$ TeV are shown in the left inset as lines with error bands (the total error is obtained by adding all given uncertainties, i.e. statistical, systematic, luminosity, and beam uncertainties, in quadrature). 
In the central inset, we show the publicly available combinations for the ATLAS and CMS measurements~\cite{ATLAS:2022aof} of $178.5 \pm 4.7$ pb at collision energy $\sqrt{S}=7$ TeV, and of $243.3^{+6.0}_{-5.9}$ pb at collision energy of 8 TeV. The error bands are obtained, as before, by adding the uncertainties in quadrature. We observe that the experimental uncertainties at these two energies are smaller that the theoretical ones at aN$^3$LO QCD+EW.
In the right inset, we show the result of $829 \pm 15$ pb from ATLAS~\cite{ATLAS:2023gsl} and of $791 \pm 25$ pb from CMS~\cite{CMS:2021vhb} at $\sqrt{S}=13$ TeV. Here, the uncertainty of the theory prediction at aN$^3$LO QCD+EW is comparable with the experimental ones. However, the recent ATLAS result~\cite{ATLAS:2023gsl} has a smaller uncertainty due to a better control on the systematic errors.

In the lower plot of Fig.~\ref{xsections}, we show the most recent available measurements at $\sqrt{S}=13.6$ TeV of $859 \pm 29$ pb at ATLAS~\cite{ATLAS:2023sjw} and of $882 \pm 30$ pb at CMS~\cite{CMS:2023qyl} together with theory predictions calculated at $\sqrt{S}=13.6$ as well as at 14 TeV. The theoretical uncertainty at aN$^3$LO QCD+EW is comparable with the experimental ones at 13.6 TeV.

The impact of the aN$^3$LO corrections in the hard scattering consistently increases the total inclusive cross section at all collision energies. 
When the aN$^3$LO partonic cross section is convoluted with the MSHT20 aN$^3$LO pdf, this increment is mitigated in part by the approximate N$^3$LO pdf evolution, and in part by a softer gluon at large $x$ resulting from the global QCD analysis.

\section{Top-quark $p_T$ distributions at 13 TeV}
\label{pTt distributions at 13 TeV}

In this section we provide theoretical predictions for top-quark transverse momentum ($p_T$) differential distributions in top-quark pair production up to aN$^3$LO in QCD at the LHC with a collision energy of 13 TeV. 
We include electroweak corrections at NLO in the electroweak coupling constant $\alpha$: we consider terms of order ${\cal O}(\alpha_s^2\alpha)$ and subleading contributions of order  ${\cal O}(\alpha_s \alpha^2)$ and ${\cal O}(\alpha^3)$. 
The combined QCD$\times$EW theory predictions, which also include order ${\cal O}(\alpha_s^3 \alpha)$ terms, are obtained using the multiplicative method as discussed in Ref.~\cite{Czakon:2017wor}. The results are presented using the binning of Ref.~\cite{CMS:2018adi} (six $p_T$ bins) and describe a top-quark transverse momentum spectrum up to 550 GeV.
Theory predictions are shown in Sec.~\ref{theory-predictions} and they are compared to single-differential cross section measurements from the LHC at 13 TeV collision energy \cite{CMS:2018adi,ATLAS:2019hxz} in Sec.~\ref{LHC-13TeV-comp}. A brief comparison with previous theoretical top-$p_T$ predictions is given in Sec. \ref{previous-top-pT}.

\subsection{Theoretical predictions for $p_T$ distributions}
\label{theory-predictions}

Predictions for the top-quark $p_T$ distribution through NNLO QCD are computed with the publicly available computer program \texttt{MATRIXv2.1.0}~\cite{Grazzini:2017mhc}, which is based on previous works~\cite{Catani:2009sm,Catani:2007vq} 
 and uses tools for amplitude computation~\cite{Cascioli:2011va,Denner:2016kdg,Buccioni:2019sur}. These results are cross-checked at LO and NLO using in-house codes as well as \texttt{MadGraph5 aMC@NLO} \cite{MG5} and \texttt{fastNLO} \cite{Kluge:2006xs,Wobisch:2011ij,Britzger:2012bs,Britzger:2015ksm,Czakon:2017dip}.
 
The prediction at NNLO QCD$\times$EW is obtained by multiplying the NNLO QCD distributions with proper EW $K$-factors that are 
obtained by using \texttt{fastNLO} tables with LUXQED17 NNLO pdf~\cite{Manohar:2017eqh}. These predictions are publicly available at~\cite{fastNLOew}.
We checked that the EW $K$-factors do not change with scale variation and pdf, and we therefore assume they are scale and pdf independent, at least for the precision given in our results. 

The aN$^3$LO QCD corrections are calculated using the theoretical work in \cite{NKan3lo2}, and they are added to the NNLO QCD prediction to derive the aN$^3$LO QCD top-quark $p_T$ distribution. New predictions at aN$^3$LO QCD$\times$EW accuracy are obtained by multiplying the aN$^3$LO QCD result by EW $K$-factors.

\begin{figure}[htbp]
\begin{center}
\includegraphics[width=135mm]{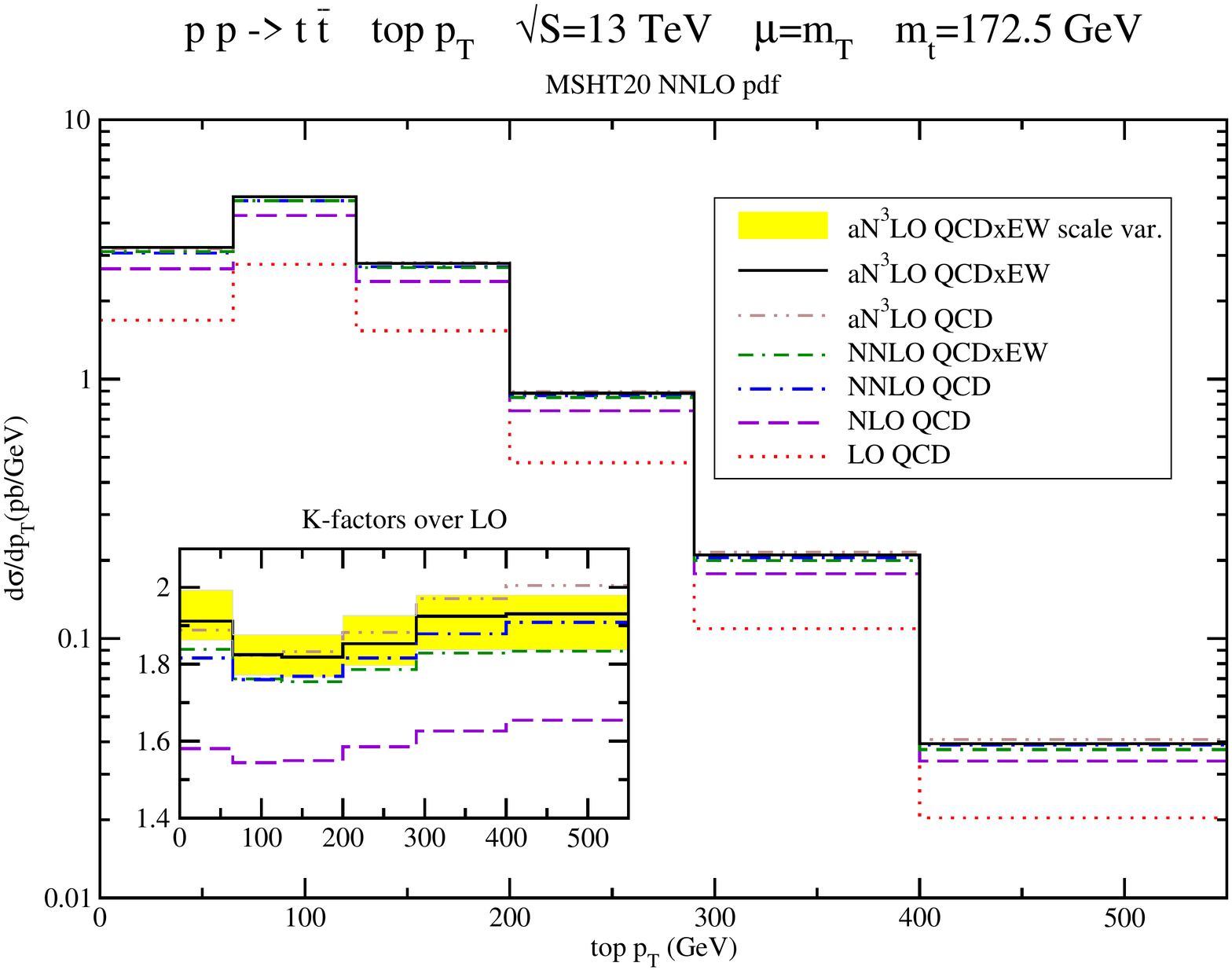}
\end{center}
\vspace{5mm}
\begin{center}
\includegraphics[width=135mm]{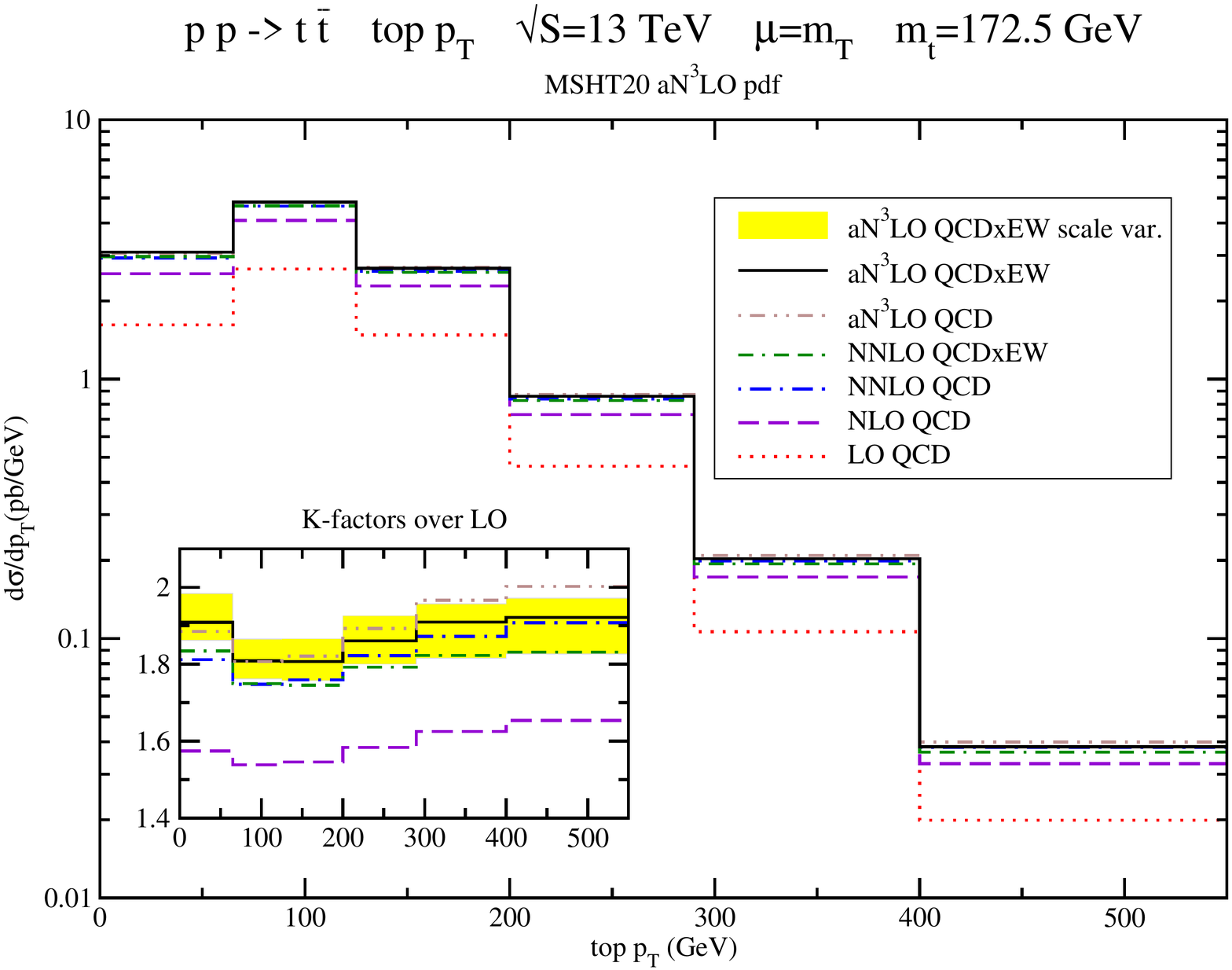}
\caption{Top-quark $p_T$ distribution at different perturbative orders with MSHT20 NNLO pdf (upper plot) and MSHT20 aN$^3$LO pdf (lower plot), and $\mu=m_T$. 
Scale variations $m_T/2\leq \mu\leq 2 m_T$ for the aN$^3$LO QCD$\times$EW prediction are represented by the yellow band in each plot.}
\label{pt_msht_theory_1}
\end{center}
\end{figure}

\begin{figure}[htbp]
\begin{center}
\includegraphics[width=135mm]{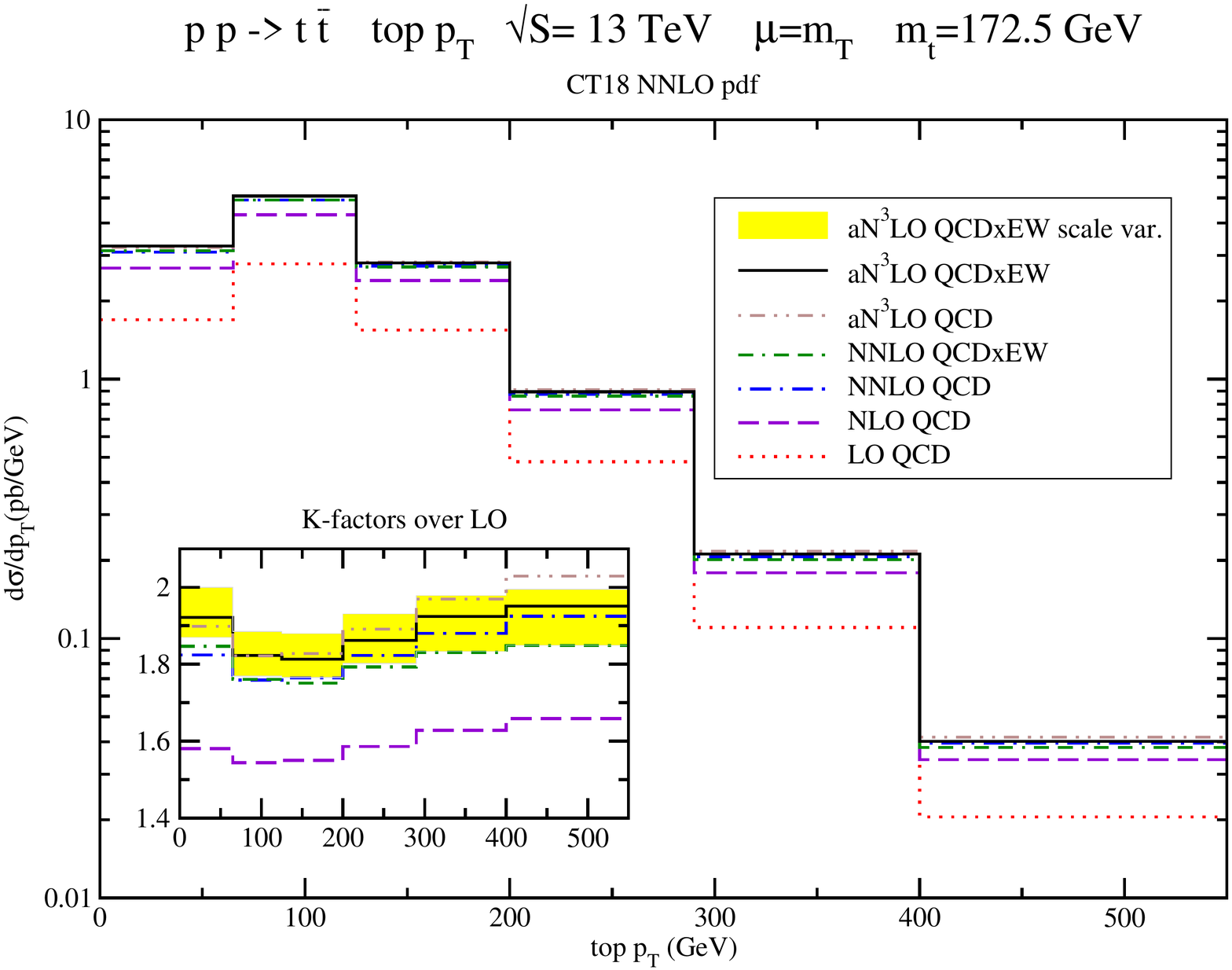}
\end{center}
\vspace{5mm}
\begin{center}
\includegraphics[width=135mm]{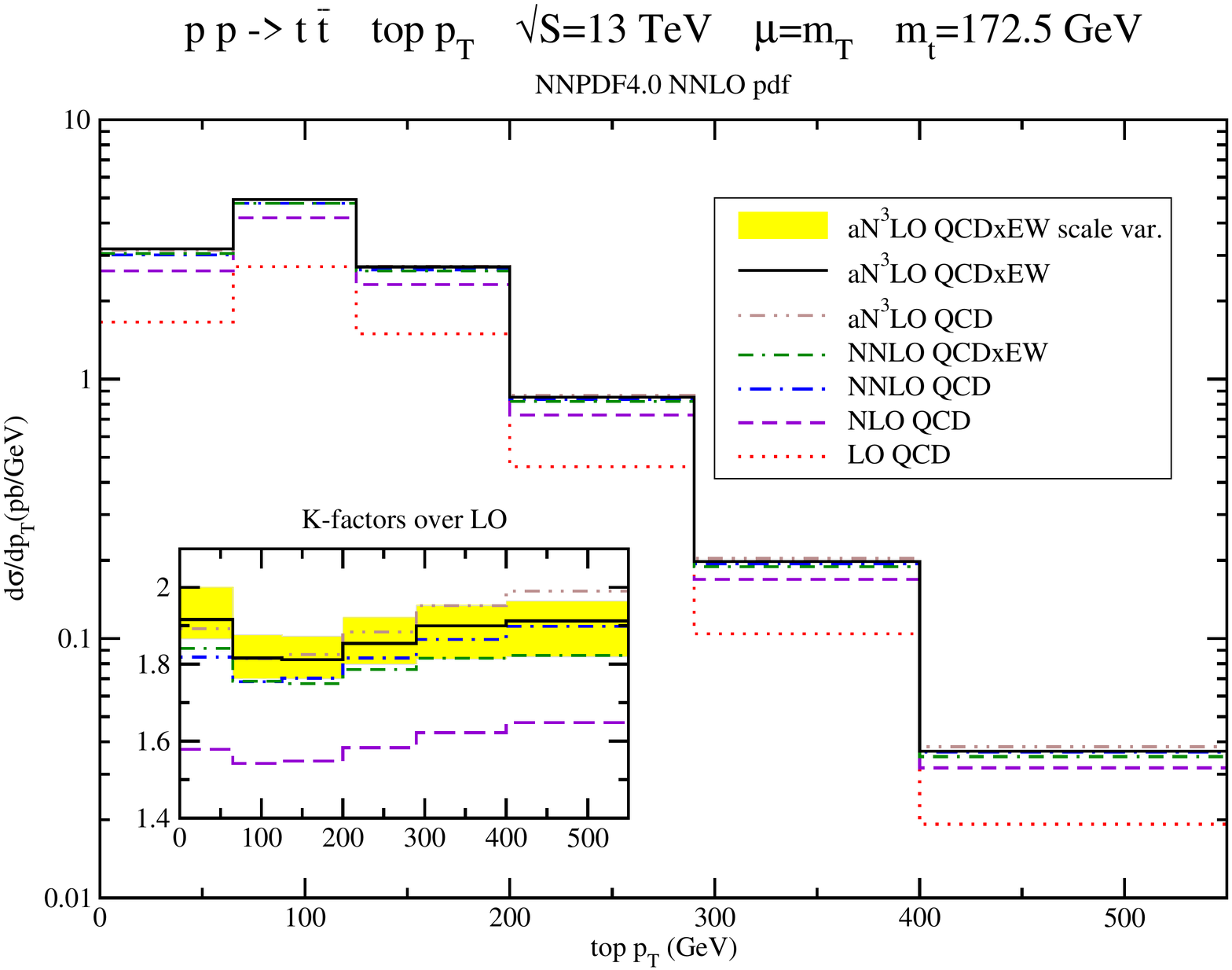}
\caption{
Top-quark $p_T$ distribution at different perturbative orders with CT18 NNLO pdf (upper plot) and NNPDF4.0 NNLO pdf (lower plot), and $\mu=m_T$. 
Scale variations $m_T/2\leq \mu\leq 2 m_T$ for the aN$^3$LO QCD$\times$EW prediction are represented by the yellow band in each plot.}
\label{pt_ct_nn_theory_1}
\end{center}
\end{figure}

In Figs.~\ref{pt_msht_theory_1} and ~\ref{pt_ct_nn_theory_1} we present the results for four different pdf sets, MSHT20 NNLO~\cite{MSHT20}, MSHT20 aN$^3$LO~\cite{MSHT20a}, CT18 NNLO~\cite{CT18}, and NNPDF4.0 NNLO~\cite{NNPDF}. These pdf sets are used in the calculation of each perturbative order as we are interested in the growth of the perturbative series. For the top-quark $p_T$ distribution the central scale is set to $\mu=m_T$, where $m_T=(p_T^2+m_t^2)^{1/2}$ is the top-quark transverse mass. Scale uncertainties are estimated by varying the common scale $\mu$ in the range $m_T/2\leq \mu\leq 2 m_T$. We also checked for the $p_T$ distributions that uncertainties obtained from the envelope of a 7-point scale variation, where $\mu_F$ and $\mu_R$ are varied independently, are basically the same as the ones obtained by performing a simpler 3-point scale variation. Uncertainties induced by pdf errors will be shown in Sec.~\ref{LHC-13TeV-comp} when comparing with data. 

In Fig.~\ref{pt_msht_theory_1} we show theoretical predictions (central value) for the top-quark $p_T$ distribution at $\sqrt{S}=13$ TeV collision energy obtained with MSHT20 NNLO pdf (upper plot) and MSHT20 aN$^3$LO pdf (lower plot), respectively. These are shown for four perturbative orders in QCD, i.e., LO, NLO, NNLO, and aN$^3$LO, and we also include the NNLO QCD$\times$EW, and aN$^3$LO QCD$\times$EW theory predictions. $K$-factors over the LO QCD results are shown in each inset plot where the yellow band represents the scale uncertainty of the aN$^3$LO QCD$\times$EW prediction. 

In  Fig.~\ref{pt_ct_nn_theory_1}, we show the corresponding theoretical predictions for the top-quark $p_T$ distribution at $\sqrt{S}=13$ TeV with CT18 NNLO pdf (upper plot) and NNPDF4.0 NNLO pdf (lower plot), respectively.

\begin{table}[htb!]
\begin{center}
\begin{tabular}{|c|c|c|c|c|} \hline
\multicolumn{4}{|c|}{Top-quark $p_T$ distribution at 13 TeV with MSHT20 NNLO pdf } \\ \hline
$d\sigma/dp_T$ in pb/GeV & aN$^3$LO QCD & aN$^3$LO QCD$\times$EW & aN$^3$LO QCD /  \\ &  &  &  NNLO  QCD \\ \hline
$0 < p_T < 65$ GeV & $3.18^{+0.14}_{-0.08}{}^{+0.06}_{-0.04}$ & $3.22^{+0.14}_{-0.08}{}^{+0.06}_{-0.04}$ & 1.040 \\ \hline
$65 < p_T < 125$ GeV & $5.05^{+0.14}_{-0.15}{}^{+0.09}_{-0.07}$ & $5.05^{+0.14}_{-0.15}{}^{+0.10}_{-0.07}$ & 1.036 \\ \hline
$125 < p_T < 200$ GeV & $2.81^{+0.09}_{-0.08}{}^{+0.06}_{-0.03}$ & $2.79^{+0.09}_{-0.08}{}^{+0.06}_{-0.04}$ & 1.036 \\ \hline
$200 < p_T < 290$ GeV & $0.896^{+0.036}_{-0.027}{}^{+0.022}_{-0.014}$ & $0.882^{+0.035}_{-0.027}{}^{+0.022}_{-0.013}$ & 1.037 \\ \hline
$290 < p_T < 400$ GeV & $0.215^{+0.006}_{-0.010}{}^{+0.006}_{-0.004}$ & $0.210^{+0.006}_{-0.009}{}^{+0.006}_{-0.004}$ & 1.049 \\ \hline
$400 < p_T < 550$ GeV & $0.0408^{+0.0010}_{-0.0020}{}^{+0.0014}_{-0.0009}$ & $0.0393^{+0.0010}_{-0.0019}{}^{+0.0013}_{-0.0008}$ & 1.050 \\ \hline
\end{tabular}
\caption{The top-quark $p_T$ distribution at aN$^3$LO QCD and aN$^3$LO QCD$\times$EW, with $m_t=172.5$ GeV and MSHT20 NNLO pdf.
The central results are with $\mu=m_T$ and are shown together with scale and pdf uncertainties. The last column provides the aN$^3$LO QCD over NNLO QCD $K$-factor.}
\label{tablepT1}
\end{center}
\end{table}

In Tables~\ref{tablepT1},~\ref{tablepT2},~\ref{tablepT3}, and ~\ref{tablepT4} we present numerical results for the top-quark $p_T$ distribution in six $p_T$ bins at aN$^3$LO QCD and aN$^3$LO QCD$\times$EW, using MSHT20 NNLO pdf, MSHT20 aN$^3$LO pdf, CT18 NNLO pdf, and NNPDF4.0 NNLO pdf, respectively. The second and third columns in the tables show the central results in each bin which are calculated at scale $\mu=m_T$ and reported together with scale variation $m_T/2\leq \mu\leq 2 m_T$ and pdf uncertainties. The last column in the tables provides the aN$^3$LO QCD over NNLO QCD $K$-factor in each bin. We note that aN$^3$LO QCD$\times$EW over NNLO QCD$\times$EW is equivalent to aN$^3$LO QCD over NNLO QCD because electroweak corrections have been evaluated with a $K$-factor which cancels out in the ratio.

\begin{table}[htb!]
\begin{center}
\begin{tabular}{|c|c|c|c|c|} \hline
\multicolumn{4}{|c|}{Top-quark $p_T$ distribution at 13 TeV with MSHT20 aN$^3$LO pdf } \\ \hline
$d\sigma/dp_T$ in pb/GeV & aN$^3$LO QCD & aN$^3$LO QCD$\times$EW & aN$^3$LO QCD / \\ &  &  &  NNLO  QCD \\ \hline
$0 < p_T < 65$ GeV & $3.05^{+0.12}_{-0.07}{}^{+0.06}_{-0.06}$ & $3.09^{+0.12}_{-0.08}{}^{+0.06}_{-0.06}$ & 1.041 \\ \hline
$65 < p_T < 125$ GeV & $4.81^{+0.15}_{-0.13}{}^{+0.09}_{-0.10}$ & $4.81^{+0.15}_{-0.12}{}^{+0.10}_{-0.10}$ & 1.034 \\ \hline
$125 < p_T < 200$ GeV & $2.70^{+0.09}_{-0.08}{}^{+0.05}_{-0.07}$ & $2.68^{+0.08}_{-0.08}{}^{+0.05}_{-0.07}$ & 1.035 \\ \hline
$200 < p_T < 290$ GeV & $0.873^{+0.031}_{-0.028}{}^{+0.021}_{-0.023}$ & $0.858^{+0.030}_{-0.028}{}^{+0.021}_{-0.022}$ & 1.039 \\ \hline
$290 < p_T < 400$ GeV & $0.209^{+0.005}_{-0.010}{}^{+0.006}_{-0.006}$ & $0.203^{+0.005}_{-0.010}{}^{+0.006}_{-0.006}$ & 1.050 \\ \hline
$400 < p_T < 550$ GeV & $0.0399^{+0.0010}_{-0.0020}{}^{+0.0012}_{-0.0013}$ & $0.0383^{+0.0010}_{-0.0019}{}^{+0.0012}_{-0.0013}$ & 1.050 \\ \hline
\end{tabular}
\caption{The top-quark $p_T$ distribution at aN$^3$LO QCD and aN$^3$LO QCD$\times$EW, with $m_t=172.5$ GeV and MSHT20 aN$^3$LO pdf.
The central results are with $\mu=m_T$ and are shown together with scale and pdf uncertainties. The last column provides the aN$^3$LO QCD over NNLO QCD $K$-factor.}
\label{tablepT2}
\end{center}
\end{table}

\begin{table}[htb!]
\begin{center}
\begin{tabular}{|c|c|c|c|c|} \hline
\multicolumn{4}{|c|}{Top-quark $p_T$ distribution at 13 TeV with CT18 NNLO pdf } \\ \hline
$d\sigma/dp_T$ in pb/GeV & aN$^3$LO QCD & aN$^3$LO QCD$\times$EW & aN$^3$LO QCD / \\ &  &  &  NNLO  QCD \\ \hline
$0 < p_T < 65$ GeV & $3.22^{+0.13}_{-0.09}{}^{+0.06}_{-0.06}$ & $3.26^{+0.13}_{-0.09}{}^{+0.06}_{-0.06}$ & 1.041 \\ \hline
$65 < p_T < 125$ GeV & $5.07^{+0.17}_{-0.15}{}^{+0.10}_{-0.10}$ & $5.08^{+0.17}_{-0.15}{}^{+0.10}_{-0.10}$ & 1.036 \\ \hline
$125 < p_T < 200$ GeV & $2.83^{+0.10}_{-0.08}{}^{+0.06}_{-0.06}$ & $2.81^{+0.10}_{-0.08}{}^{+0.06}_{-0.06}$ & 1.036 \\ \hline
$200 < p_T < 290$ GeV & $0.908^{+0.034}_{-0.029}{}^{+0.027}_{-0.020}$ & $0.894^{+0.033}_{-0.029}{}^{+0.026}_{-0.020}$ & 1.038  \\ \hline
$290 < p_T < 400$ GeV & $0.217^{+0.006}_{-0.010}{}^{+0.009}_{-0.006}$ & $0.212^{+0.006}_{-0.010}{}^{+0.008}_{-0.005}$ & 1.047 \\ \hline
$400 < p_T < 550$ GeV & $0.0417^{+0.0009}_{-0.0022}{}^{+0.0022}_{-0.0013}$ & $0.0401^{+0.0009}_{-0.0021}{}^{+0.0021}_{-0.0013}$ & 1.054 \\ \hline
\end{tabular}
\caption{The top-quark $p_T$ distribution at aN$^3$LO QCD and aN$^3$LO QCD$\times$EW, with $m_t=172.5$ GeV and CT18 NNLO pdf.
The central results are with $\mu=m_T$ and are shown together with scale and pdf uncertainties. The last column provides the aN$^3$LO QCD over NNLO QCD $K$-factor.}
\label{tablepT3}
\end{center}
\end{table}

\begin{table}[htb!]
\begin{center}
\begin{tabular}{|c|c|c|c|c|} \hline
\multicolumn{4}{|c|}{Top-quark $p_T$ distribution at 13 TeV with NNPDF4.0 NNLO pdf } \\ \hline
$d\sigma/dp_T$ in pb/GeV & aN$^3$LO QCD & aN$^3$LO QCD$\times$EW & aN$^3$LO QCD / \\ &  &  &  NNLO QCD \\ \hline
$0 < p_T < 65$ GeV & $3.14^{+0.13}_{-0.09}{}^{+0.01}_{-0.02}$ & $3.18^{+0.13}_{-0.09}{}^{+0.01}_{-0.02}$ & 1.041 \\ \hline
$65 < p_T < 125$ GeV & $4.92^{+0.16}_{-0.15}{}^{+0.02}_{-0.03}$ & $4.92^{+0.17}_{-0.14}{}^{+0.03}_{-0.02}$ & 1.034 \\ \hline
$125 < p_T < 200$ GeV & $2.73^{+0.09}_{-0.08}{}^{+0.01}_{-0.02}$ & $2.71^{+0.09}_{-0.08}{}^{+0.01}_{-0.02}$ & 1.035 \\ \hline
$200 < p_T < 290$ GeV & $0.865^{+0.032}_{-0.025}{}^{+0.006}_{-0.006}$ & $0.851^{+0.031}_{-0.025}{}^{+0.006}_{-0.006}$ & 1.038 \\ \hline
$290 < p_T < 400$ GeV & $0.204^{+0.005}_{-0.010}{}^{+0.001}_{-0.002}$ & $0.198^{+0.006}_{-0.009}{}^{+0.002}_{-0.002}$ & 1.047 \\ \hline
$400 < p_T < 550$ GeV & $0.0383^{+0.0010}_{-0.0019}{}^{+0.0004}_{-0.0004}$ & $0.0368^{+0.0010}_{-0.0018}{}^{+0.0004}_{-0.0004}$ & 1.048 \\ \hline
\end{tabular}
\caption{The top-quark $p_T$ distribution at aN$^3$LO QCD and aN$^3$LO QCD$\times$EW, with $m_t=172.5$ GeV and NNPDF4.0 NNLO pdf.
The central results are with $\mu=m_T$ and are shown together with scale and pdf uncertainties. The last column provides the aN$^3$LO QCD over NNLO QCD $K$-factor.}
\label{tablepT4}
\end{center}
\end{table}

In general, EW corrections provide a positive contribution to the cross section at lower $p_T$ while the contributions are increasingly negative at higher $p_T$ (making the effect to the total cross section very small due to this partial cancellation).

Moreover, we note that the CT18 NNLO pdf provides cross sections values in every bin which are systematically bigger than the ones obtained from the other pdf sets, while MSHT20 aN$^3$LO and NNPDF4.0 NNLO pdf provide the smallest ones, depending on the considered bin. Nonetheless, the $K$-factors are very similar for all four pdf sets used in this calculation. The $p_T$ distributions obtained with the CT18 NNLO, MSHT20 NNLO, and MSHT20 aN$^3$LO pdf have similar and more conservative pdf uncertainties as compared to those from NNPDF4.0 NNLO in the considered $p_T$ bins.

\subsection{Comparison with 13 TeV LHC top-$p_T$ data}
\label{LHC-13TeV-comp}

In this section, we compare theoretical predictions for the top-quark $p_T$ distributions at NNLO QCD$\times$EW and aN$^3$LO QCD$\times$EW 
to the $\sqrt{S}=13$ TeV high-precision measurements from CMS~\cite{CMS:2018adi} in the dilepton channel with 35.9 fb$^{-1}$ of integrated luminosity, and from  ATLAS~\cite{ATLAS:2019hxz} in the lepton+jets channel with 36 fb$^{-1}$ of integrated luminosity. As before, we use $\mu=m_T$ for the central prediction results. 
The differential cross section uncertainties from pdf errors are computed by using \texttt{fastNLO} tables for the NNLO predictions~\cite{Czakon:2017dip} and the \texttt{fastNLO-toolkit-v2.5.0}~\cite{Kluge:2006xs,Wobisch:2011ij,Britzger:2012bs,Britzger:2015ksm}.

\begin{figure}[htbp]
\begin{center}
\includegraphics[width=140mm]{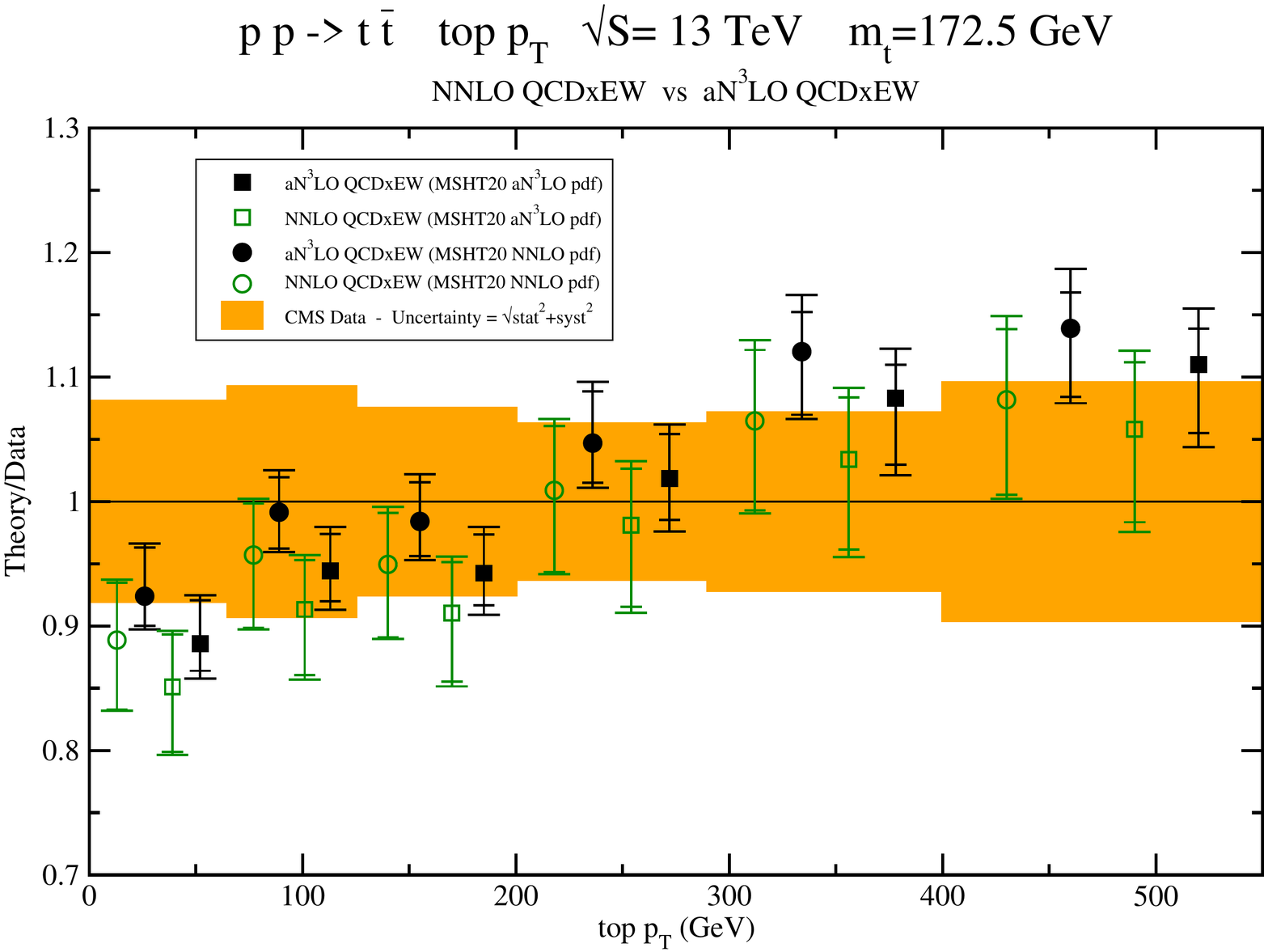}
\includegraphics[width=140mm]{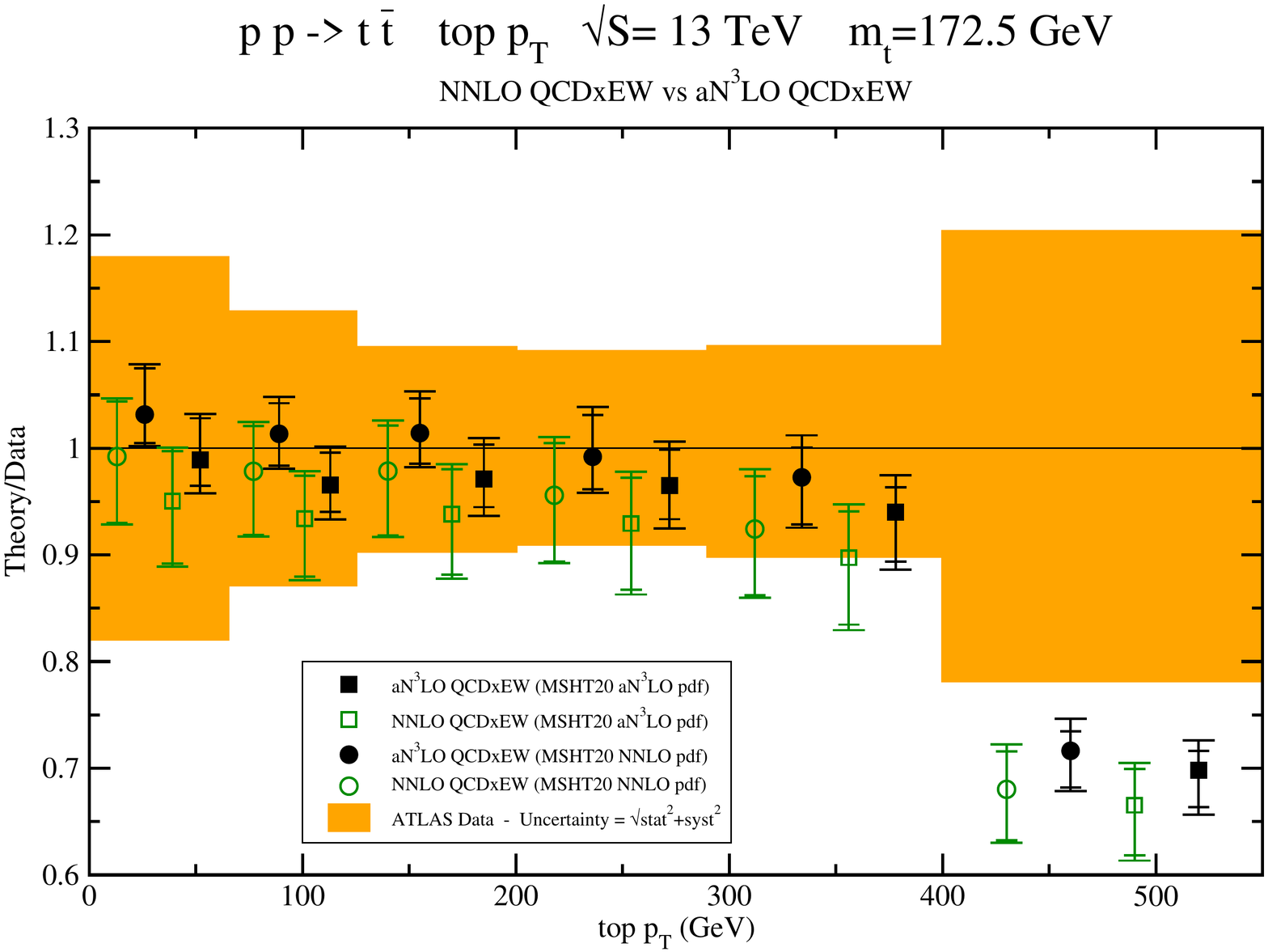}
\caption{Comparison of NNLO QCD$\times$EW and aN$^3$LO QCD$\times$EW theory predictions using MSHT20 NNLO and aN$^3$LO pdf with CMS (upper plot) and ATLAS (lower plot) top-quark transverse momentum  data. The orange band represents the sum of statistical and systematic experimental uncertainties added in quadrature. Inner (outer) bars represent scale (scale plus pdf) theoretical uncertainties.}
\label{pt_msht_data_1_bar}
\end{center}
\end{figure}

\begin{figure}[htbp]
\begin{center}
\includegraphics[width=140mm]{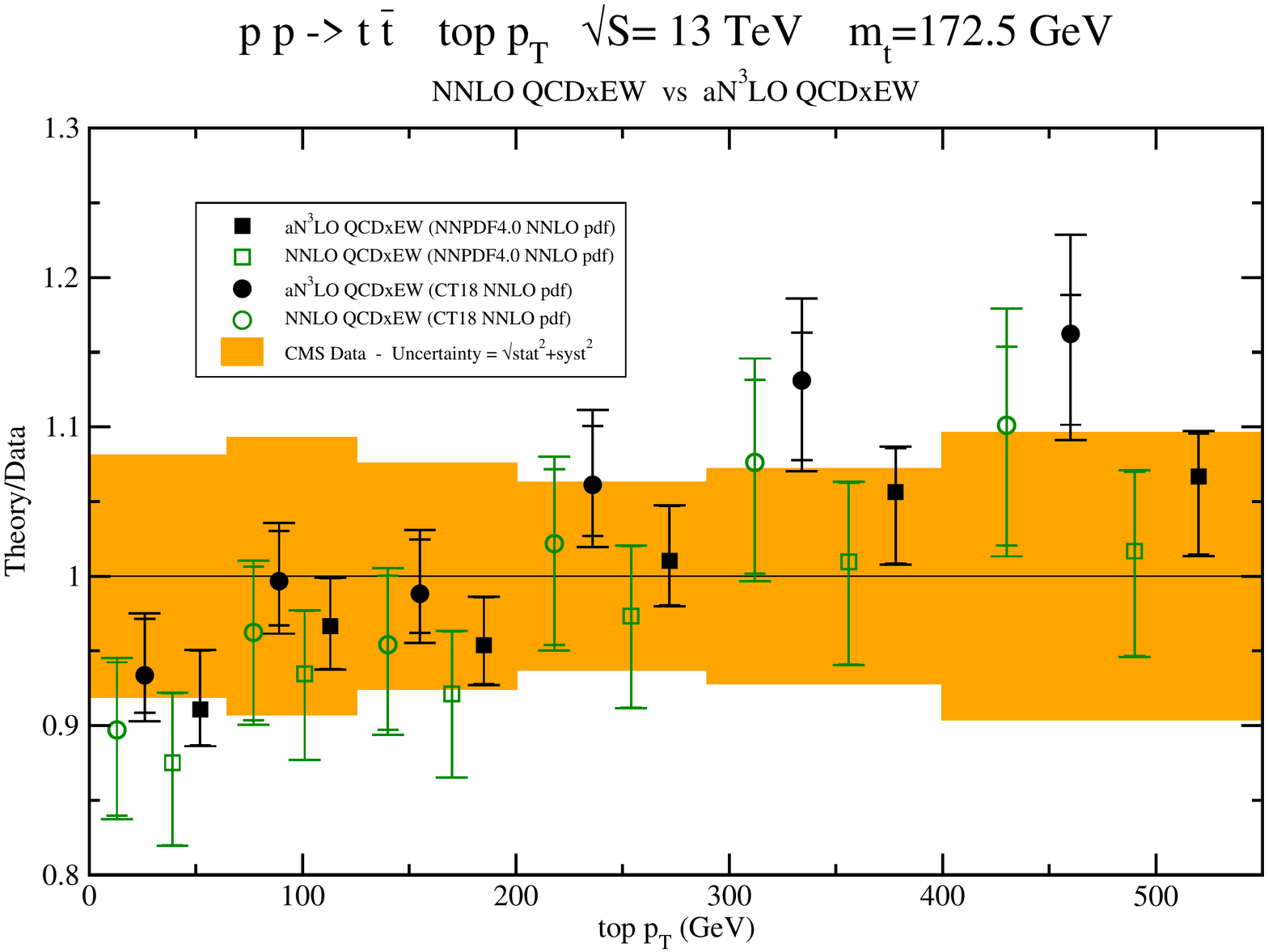}
\includegraphics[width=140mm]{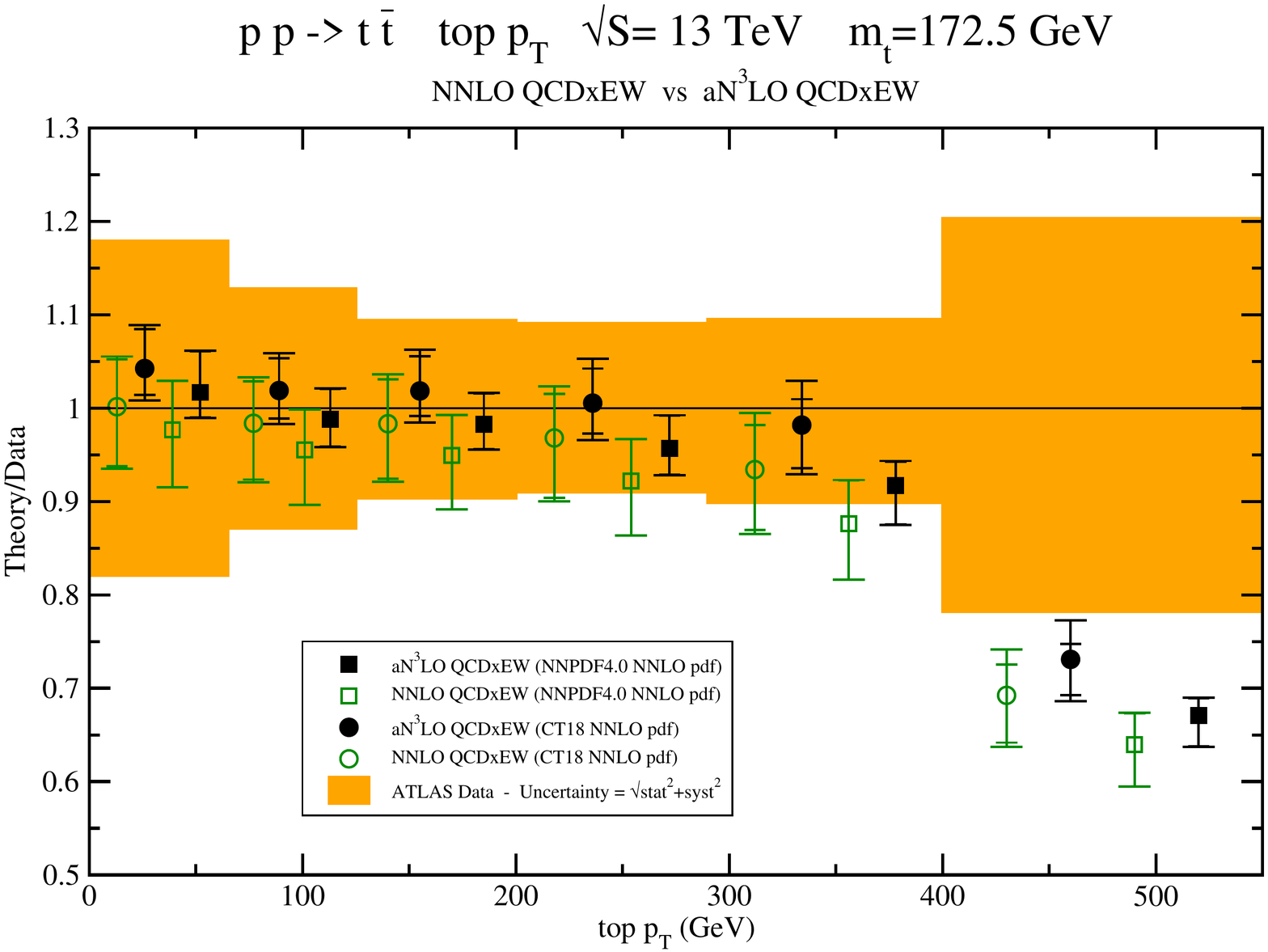}
\caption{Comparison of NNLO QCD$\times$EW and aN$^3$LO QCD$\times$EW theory predictions using CT18 and NNPDF4.0 NNLO pdf with CMS (upper plot) and ATLAS (lower plot) top-quark transverse momentum data. The orange band represents the sum of statistical and systematic experimental uncertainties added in quadrature. Inner (outer) bars represent scale (scale plus pdf) theoretical uncertainties.}
\label{pt_ct_data_1_bar}
\end{center}
\end{figure}
In Fig.~\ref{pt_msht_data_1_bar}, the top-quark $p_T$ distributions at NNLO QCD$\times$EW and aN$^3$LO QCD$\times$EW obtained with 
MSHT20 NNLO and aN$^3$LO pdf are compared with ATLAS and CMS measurements. In the two plots we show the ratio of our theoretical predictions to the data together with scale and 68\% C.L. pdf uncertainties.  
The orange band represents the experimental statistical and systematical uncertainties added in quadrature. In analogy, in Fig.~\ref{pt_ct_data_1_bar}  we compare CMS and ATLAS data with the theoretical prediction for the differential distribution in top-quark transverse momentum at NNLO QCD$\times$EW and aN$^3$LO QCD$\times$EW obtained using CT18 and NNPDF4.0 NNLO pdf. We observe that there are differences in the measured values of the top-$p_T$ distribution at CMS and ATLAS, especially at large $p_T$ where the two measurements pull in opposite directions.

\begin{table}[!ht]
\begin{center}
\begin{tabular}{|c|c|c|c|c|}
\hline
  pdf & NNLO QCD & NNLO QCD   & aN$^3$LO QCD & aN$^3$LO QCD\\
        &                     &    $\times$EW &                           & $\times$EW\\
\hline
\hline
MSHT20 NNLO    & 2.57 & 1.58 & 3.27 & 2.15\\
MSHT20 aN$^3$LO  & 2.76 & 1.80 &3.42 & 2.20\\
CT18 NNLO         & 2.86 & 1.79& 3.68 & 2.44\\
NNPDF4.0 NNLO & 1.56 & 0.91& 1.92 & 1.09\\ 
\hline
\end{tabular}
\end{center}
\caption{Summary of the $\chi^2/N_{pt}$ for the top-quark $p_T$ distributions at CMS.}
\label{pT-chi2-CMS}
\end{table}

The quality of agreement of the theoretical predictions with the $p_T$ distribution at CMS is summarized in Table~\ref{pT-chi2-CMS} where $\chi^2/N_{pt}$ values are reported for the theory predictions calculated at NNLO QCD and aN$^3$LO QCD, with and without EW corrections, and obtained using MSHT20 NNLO, MSHT20 aN$^3$LO, CT18 NNLO, and NNPDF4.0 NNLO pdf.  

The ATLAS and CMS collaborations at the LHC have published measurements where experimental uncertainties are given in terms of either the covariance matrix or the nuisance parameters representation. Therefore, depending on the information relative to the statistical, uncorrelated, and correlated systematical errors provided with the measurements, the $\chi^2$ function must be consistently computed using one or the other representation. 

The statistical and correlated systematic uncertainties released by the CMS collaboration are given in terms of the covariance matrix representation.   
Therefore, the $\chi^2$ definition used to obtain the results in Table~\ref{pT-chi2-CMS} is 
\begin{equation}\label{chi2-CMS-formula}
\chi^2 = \sum_{i,j=1}^{N_{pt}}\left(D_i - T_i\right)  \left(\textrm{cov}^{-1}\right)_{ij}  \left(D_j - T_j\right)\,,
\end{equation}
where $N_{pt}$ is the number of data points, $\textrm{cov}$ is the covariance matrix, $D_i$ is the $i$-th data point, and $T_i$ is the corresponding theory prediction.

\begin{table}[!ht]
\begin{center}
\begin{tabular}{|c|c|c|c|c|}
\hline
  pdf & NNLO QCD & NNLO QCD   & aN$^3$LO QCD & aN$^3$LO QCD\\
        &                     &    $\times$EW &                           & $\times$EW\\
\hline
\hline
MSHT20 NNLO    & 1.07 & 1.27 & 1.40 & 1.48\\
MSHT20 aN$^3$LO  & 1.05 & 1.22 & 1.42 & 1.43\\
CT18 NNLO         & 1.17 & 1.30& 1.53 & 1.57\\
NNPDF4.0 NNLO & 1.18 & 1.58& 1.32 & 1.62\\ 
\hline
\end{tabular}
\end{center}
\caption{Summary of the $\chi^2/N_{pt}$ for the top-quark $p_T$ distributions at ATLAS.}
\label{pT-chi2-ATL}
\end{table}

The corresponding summary for the ATLAS top-$p_T$ measurements is in Table~\ref{pT-chi2-ATL}. In the ATLAS case, correlated systematic uncertainties are given in terms of nuisance parameters, and the general $\chi^2$ definition adopted in this case is
\begin{equation}\label{chi2-ATLAS-formula}
\chi^2(\lambda) = \sum_{i=1}^{N_{pt}} \frac{1}{s_i^2}\left(D_i - T_i - \sum_{\alpha=1}^{N_\lambda} \lambda_\alpha \beta_{i\alpha}\right)^2 +  \sum_{\alpha=1}^{N_\lambda} \lambda_\alpha^2\,,
\end{equation}
where  $s_i = \sqrt{s_{i,\textrm{\tiny stat}}^2 + s_{i,\textrm{\tiny uncorsys}}^2}$, with $s_{i,\textrm{\tiny stat}}$ the uncorrelated statistical error and $s_{i,\textrm{\tiny uncorsys}}$ the uncorrelated systematical error,
$N_{\lambda}$ is the number of correlated systematic uncertainties, $\lambda_\alpha$ are the nuisance parameters, and $\beta_{i\alpha}$ is the correlation matrix. 
The up and down shifts in the correlated systematic uncertainties released by the ATLAS collaboration are very asymmetric.
Therefore, in the construction of the $\beta_{i\alpha}$ matrix to calculate the $\chi^2$ in terms of nuisance parameters, we considered symmetric correlated shifts using the downward excursions. This choice gives the most conservative estimate for the $\chi^2$. The same approach is used in the determination of the $\chi^2$ for the rapidity distributions in Sec.~\ref{Y-chi2}.

Overall, the theory predictions obtained by using the four pdf sets considered in this study give similar description of the ATLAS data, while the CMS data are better described by NNPDF4.0 as compared to the other pdf sets.
We note in fact, that the NNPDF4.0 global analysis includes 13 TeV single-differential cross section measurements at CMS in the lepton+jets~\cite{CMS:2018htd} and dilepton channel~\cite{CMS:2018adi}.

We also note that the inclusion of EW corrections in either NNLO or aN$^3$LO results decreases the value of $\chi^2$ for the CMS data but increases it for the ATLAS data. Also, the $\chi^2$ at aN$^3$LO is higher than at NNLO.  It is difficult to draw meaningful conclusions for the $\chi^2$ for the $p_T$ distribution due to the big differences between CMS and ATLAS data at high $p_T$. In general, the differences we observe in the LHC data we considered may potentially generate opposite pulls in the gluon in global QCD analyses.
  
\subsection{Comparison with previous theoretical top-$p_T$ predictions}
\label{previous-top-pT}

Results at aN$^3$LO QCD for the top-quark $p_T$ distribution have been presented before using older pdf sets, beginning with Ref.~\cite{NKan3lo2}. Many comparisons with past top-$p_T$ data from the LHC were presented in the review paper of Ref.~\cite{NKtoprev}. Those theoretical results were presented as functions of the top-quark transverse momentum (not as bins) and, thus, could not and were not matched to the exact NNLO QCD results; rather, the aNNLO and aN$^3$LO soft-gluon corrections were added to the NLO result. The binned aNNLO distributions are very close to the exact NNLO ones, and hence the difference between the matched and unmatched aN$^3$LO distributions is negligible. We have checked that again for the current pdf sets, but one can also see this in past results, e.g. in Ref.~\cite{DPF2019} where the predictions are compared to the same CMS top-$p_T$ data as here.

\section{Top-quark rapidity distributions at 13 TeV}
\label{yt distributions at 13 TeV}

In this section we provide theoretical predictions for top-quark rapidity ($Y$) differential distributions in top-quark pair production up to aN$^3$LO in QCD at the LHC with a collision energy of 13 TeV. 
In analogy to the transverse momentum distribution case discussed in Sec.~\ref{pTt distributions at 13 TeV}, we include electroweak corrections at NLO in the electroweak coupling constant $\alpha$ and provide combined QCD$\times$EW theory predictions using the multiplicative method of Ref.~\cite{Czakon:2017wor}. The results are presented using ten bins of rapidity ($-2.6\leq Y\leq 2.6$) with bin size chosen according to Ref.~\cite{CMS:2018adi}.
Theory predictions are shown in Sec.~\ref{yt-theory} and they are compared to the 13 TeV differential cross section measurements at CMS~\cite{CMS:2018adi} and ATLAS~\cite{ATLAS:2019hxz} in Sec.~\ref{Y-chi2}. A brief comparison with previous theoretical top-rapidity predictions is given in Sec. \ref{previous-top-Y}.

\subsection{Theoretical predictions for rapidity distributions}
\label{yt-theory}
Theoretical predictions for the top-quark $Y$ distribution through NNLO in QCD are computed using the publicly available computer program \texttt{MATRIXv2.1.0}. 
These results are cross-checked at LO and NLO using in-house codes as well as \texttt{MadGraph5 aMC@NLO}.

The prediction at NNLO QCD$\times$EW is obtained by multiplying the NNLO QCD distributions with proper EW $K$-factors (available at~\cite{fastNLOew}) that are 
obtained by using \texttt{fastNLO} tables with LUXQED17 NNLO pdf, and are assumed to be pdf and scale independent. 

The aN$^3$LO QCD corrections are calculated using the theoretical work in \cite{NKan3lo2}, and they are added to the NNLO QCD prediction to derive the aN$^3$LO QCD top-quark rapidity distribution. New predictions at aN$^3$LO QCD$\times$EW accuracy are obtained by multiplying the aN$^3$LO QCD result by EW $K$-factors.

Results are given for four different pdf sets, MSHT20 NNLO, MSHT20 aN$^3$LO, CT18 NNLO, and NNPDF4.0 NNLO. As before, these pdf sets are used in the calculation of each perturbative order since we are interested in the growth of the perturbative series. For the top-quark $Y$ distribution the central scale is set to $\mu=m_t=172.5$ GeV. The scale uncertainties are estimated by varying the common scale $\mu$ in the range $m_t/2\leq \mu\leq 2 m_t$. We also checked for the rapidity distributions that uncertainties obtained from the envelope of a 7-point scale variation, where $\mu_F$ and $\mu_R$ are varied independently, are basically the same as the ones obtained by performing a simpler 3-point scale variation. The pdf uncertainties are not considered at this point, but they will be given in Sec.~\ref{Y-chi2}, when comparing with data.  

\begin{figure}[htbp]
\begin{center}
\includegraphics[width=135mm]{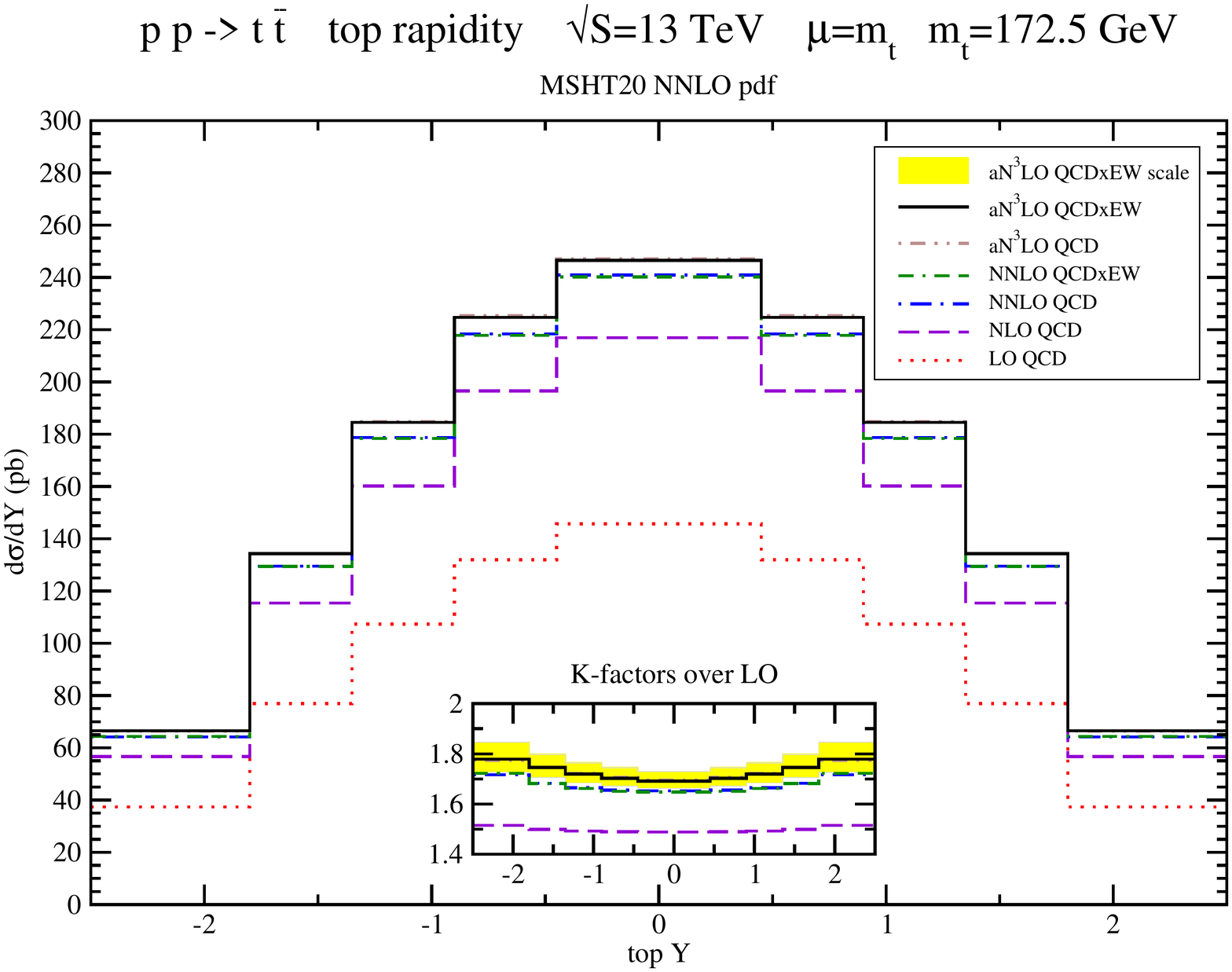}
\end{center}
\vspace{5mm}
\begin{center}
\includegraphics[width=135mm]{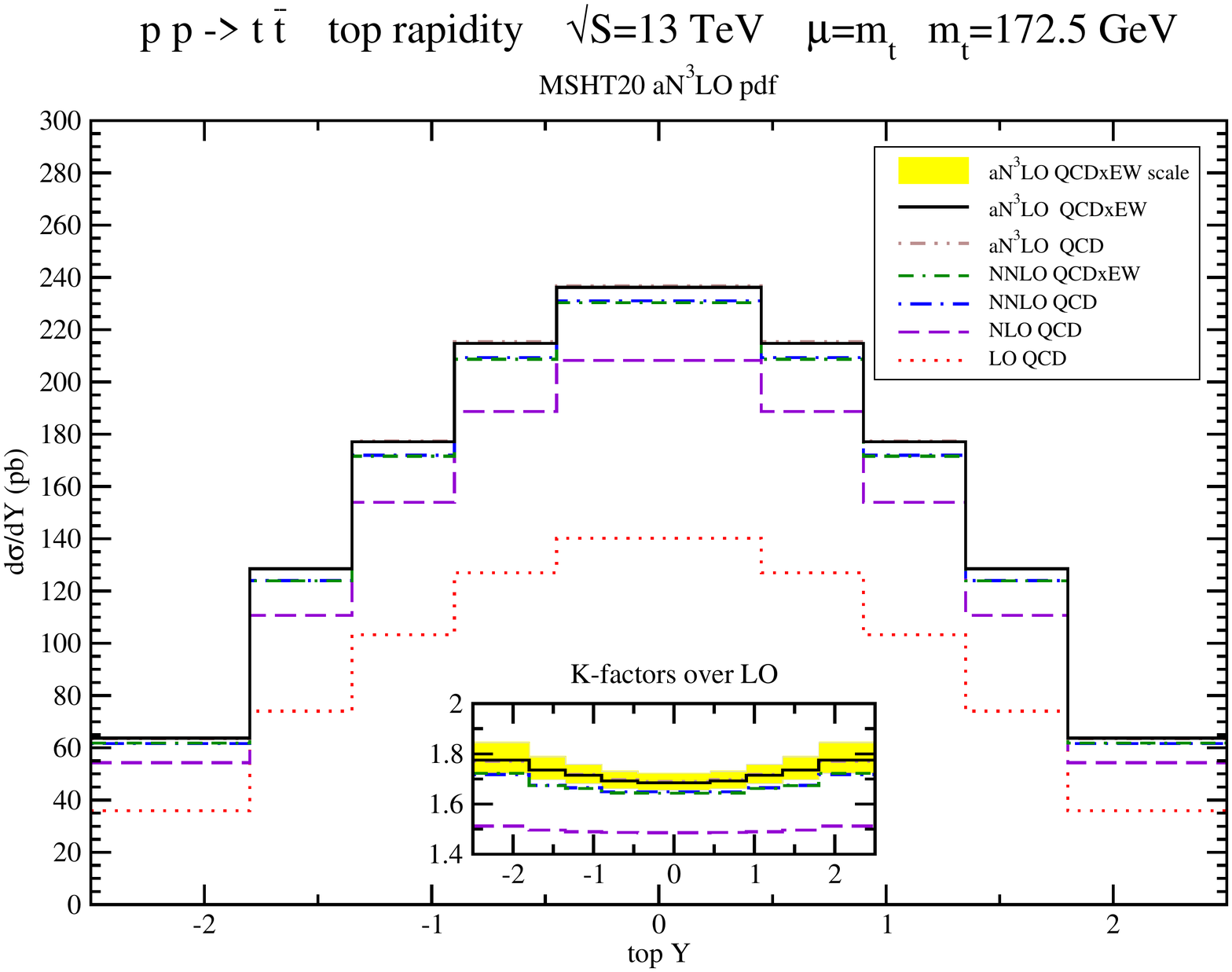}
\caption{Top-quark rapidity distribution at different perturbative orders with MSHT20 NNLO pdf (upper plot) and MSHT20 aN$^3$LO pdf (lower plot), and $\mu=m_t$. Scale variations $m_t/2\leq \mu\leq 2 m_t$ for the aN$^3$LO QCD$\times$EW prediction are represented by the yellow band in each plot.}
\label{yt_msht_theory_1}
\end{center}
\end{figure}

In the upper plot of Fig.~\ref{yt_msht_theory_1}, we show the theoretical prediction (central value) for the differential distribution in top-quark rapidity at different perturbative orders, namely LO QCD, NLO QCD, NNLO QCD, NNLO QCD$\times$EW, aN$^3$LO QCD, and aN$^3$LO QCD$\times$EW. These results have been obtained using MSHT20 NNLO pdf and $\mu=m_t$ as central scale. The $K$-factors over the LO QCD results are shown in the inset plot where the yellow band represents the scale uncertainty of the aN$^3$LO QCD$\times$EW result. The lower plot of Fig.~\ref{yt_msht_theory_1} displays the corresponding theoretical predictions obtained with MSHT20 aN$^3$LO pdf.

\begin{figure}[htbp]
\begin{center}
\includegraphics[width=135mm]{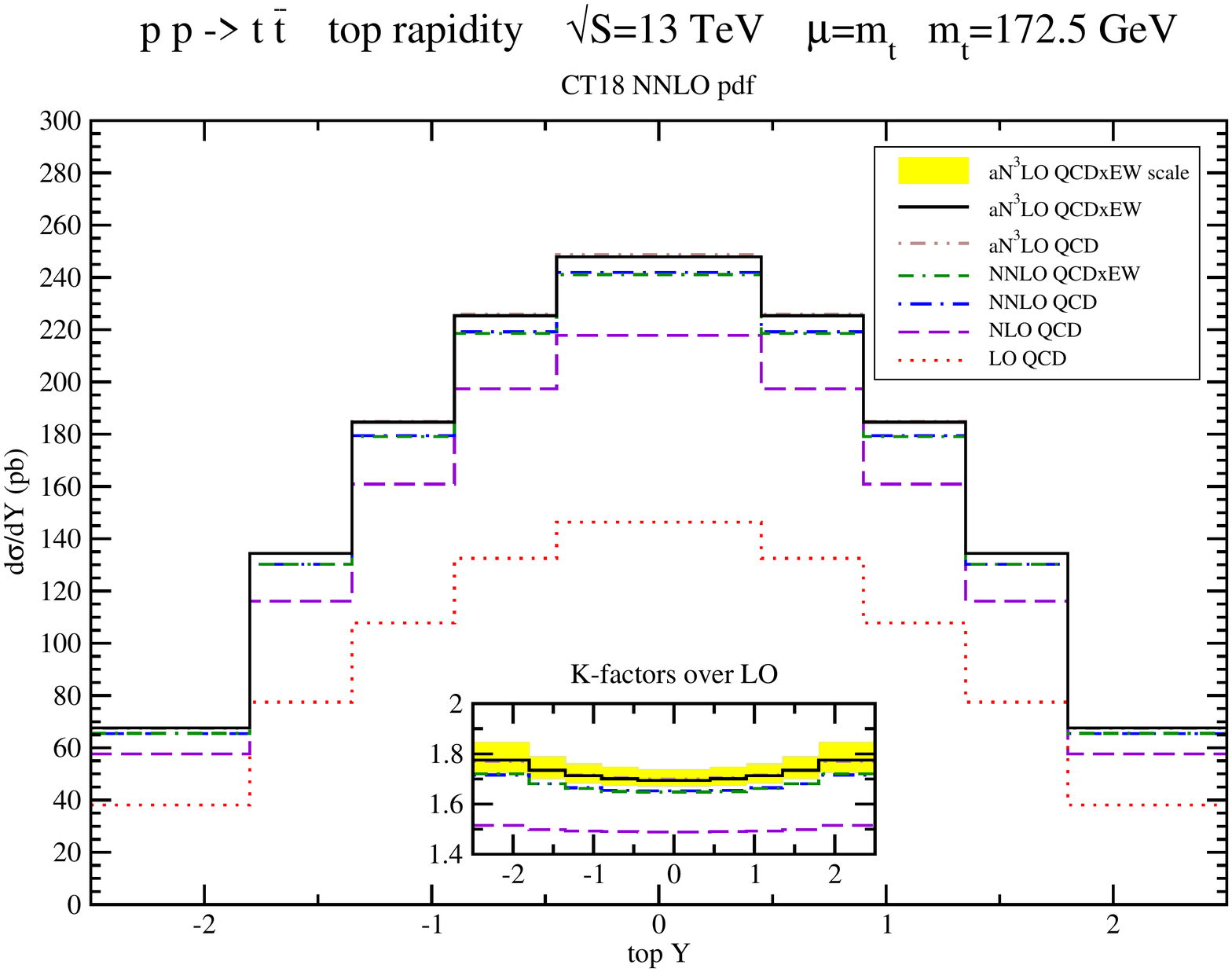}
\end{center}
\vspace{5mm}
\begin{center}
\includegraphics[width=135mm]{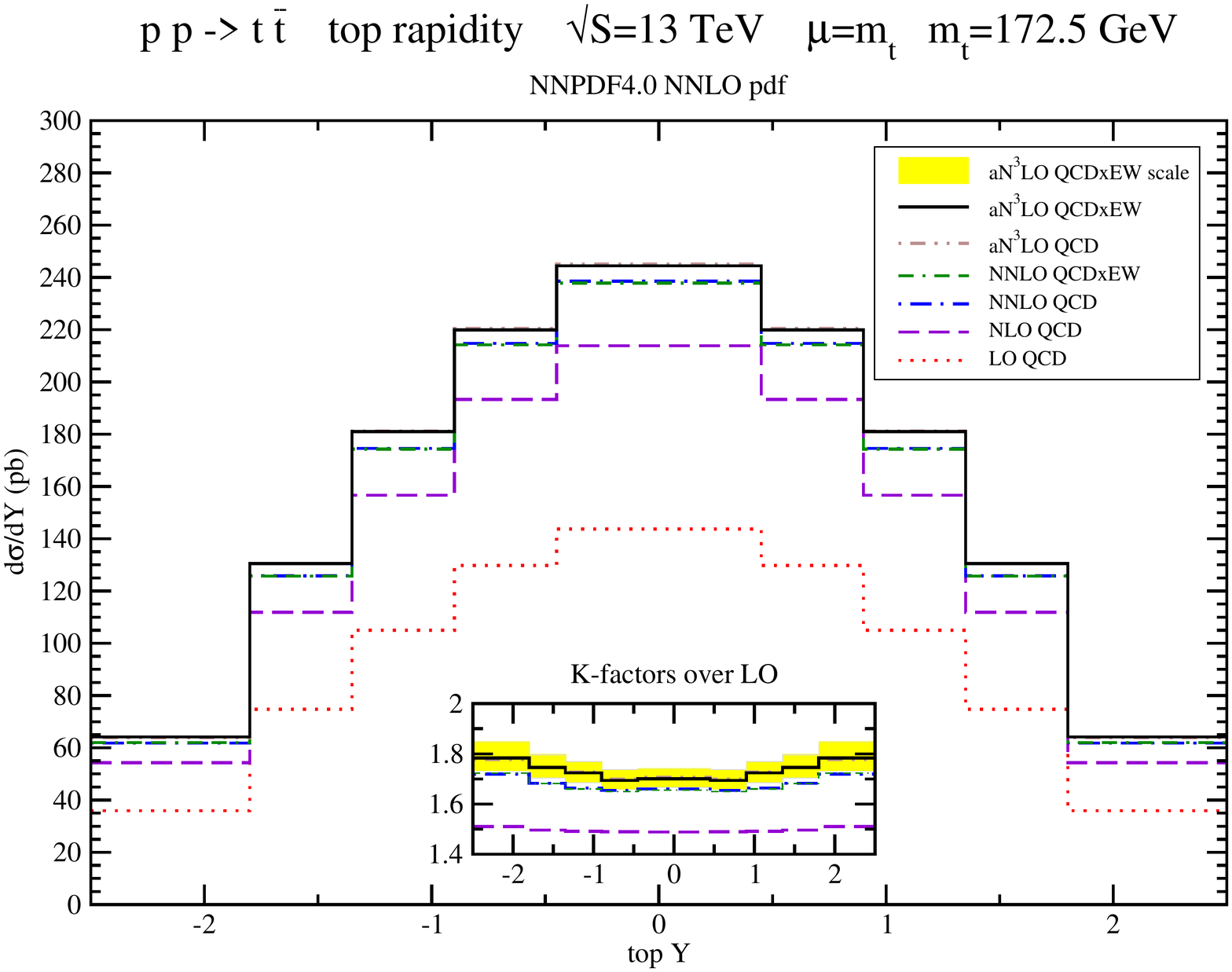}
\caption{
Top-quark rapidity distribution at different perturbative orders with CT18 NNLO pdf (upper plot) and NNPDF4.0 NNLO pdf (lower plot), and $\mu=m_t$. 
Scale variations $m_t/2\leq \mu\leq 2 m_t$ for the aN$^3$LO QCD$\times$EW prediction are represented by the yellow band in each plot.}
\label{yt_ct_nn_theory_1}
\end{center}
\end{figure}

In Fig.~\ref{yt_ct_nn_theory_1} we show the corresponding theoretical predictions for the top-quark rapidity distribution using CT18 NNLO pdf in the upper plot, and using NNPDF4.0 NNLO pdf in the lower plot.

\begin{table}[htb]
\begin{center}
\begin{tabular}{|c|c|c|c|c|} \hline
\multicolumn{4}{|c|}{Top-quark $Y$ distribution at 13 TeV with MSHT20 NNLO pdf } \\ \hline
$d\sigma/dY$ in pb & aN$^3$LO QCD & aN$^3$LO QCD$\times$EW & aN$^3$LO QCD / \\ &  &  &  NNLO  QCD \\ \hline
$0 < Y < 0.45$  & $247^{+6}_{-4}{}^{+5}_{-3}$ & $246^{+6}_{-4}{}^{+4}_{-4}$ & 1.026 \\ \hline
$0.45 < Y < 0.9$  & $225^{+6}_{-4}{}^{+4}_{-3}$ & $225^{+5}_{-4}{}^{+4}_{-3}$  & 1.032 \\ \hline
$0.9 < Y < 1.35$  & $185^{+5}_{-4}{}^{+3}_{-3}$ & $185^{+4}_{-4}{}^{+3}_{-3}$ & 1.034 \\ \hline
$1.35 < Y < 1.8$  & $134^{+4}_{-3}{}^{+3}_{-2}$ & $134^{+4}_{-3}{}^{+3}_{-2}$ & 1.037 \\ \hline
$1.8 < Y < 2.6$  & $66.3^{+2.5}_{-2.0}{}^{+1.9}_{-1.5}$ & $66.5^{+2.5}_{-2.0}{}^{+1.9}_{-1.5}$ & 1.033 \\ \hline
\end{tabular}
\caption{The top-quark rapidity distribution at aN$^3$LO QCD and aN$^3$LO QCD$\times$EW, with $m_t=172.5$ GeV and MSHT20 NNLO pdf.
The central results are with $\mu=m_t$ and are shown together with scale and pdf uncertainties. The last column provides the aN$^3$LO QCD over NNLO QCD $K$-factor.}
\label{tabley1}
\end{center}
\end{table}

In Tables~\ref{tabley1},~\ref{tabley2},~\ref{tabley3}, and ~\ref{tabley4} we present numerical results for the top-quark $Y$ distribution for five $Y$ bins (we provide results only for positive values of rapidity because the ones for negative rapidity values are symmetric), at 13 TeV collision energy, at aN$^3$LO QCD and aN$^3$LO QCD$\times$EW, using MSHT20 NNLO pdf,  MSHT20 aN$^3$LO pdf, CT18 NNLO pdf, and NNPDF4.0 NNLO pdf, respectively. In the second and third columns in the tables we show the central results in each bin which are calculated at scale $\mu=m_t$ and reported together with scale variation $m_t/2\leq \mu\leq 2 m_t$ and pdf uncertainties. The last column in the tables provides the aN$^3$LO QCD over NNLO QCD $K$-factor in each bin. Again, we note that aN$^3$LO QCD$\times$EW over NNLO QCD$\times$EW is equivalent to aN$^3$LO QCD over NNLO QCD because electroweak corrections have been evaluated with a $K$-factor which cancels out in the ratio.

\begin{table}[htb]
\begin{center}
\begin{tabular}{|c|c|c|c|c|} \hline
\multicolumn{4}{|c|}{Top-quark $Y$ distribution at 13 TeV with MSHT20 aN$^3$LO pdf } \\ \hline
$d\sigma/dY$ in pb & aN$^3$LO QCD & aN$^3$LO QCD$\times$EW & aN$^3$LO QCD / \\  &  &  &  NNLO QCD \\ \hline
$0 < Y < 0.45$  & $237^{+5}_{-4}{}^{+5}_{-5}$ & $236^{+5}_{-4}{}^{+5}_{-5}$ & 1.025 \\ \hline
$0.45 < Y < 0.9$  & $215^{+5}_{-3}{}^{+5}_{-4}$ & $215^{+5}_{-4}{}^{+4}_{-5}$ & 1.029 \\ \hline
$0.9 < Y < 1.35$  & $177^{+5}_{-3}{}^{+4}_{-3}$ & $177^{+5}_{-3}{}^{+4}_{-4}$ & 1.032 \\ \hline
$1.35 < Y < 1.8$  & $129^{+3}_{-3}{}^{+2}_{-4}$ & $129^{+3}_{-3}{}^{+2}_{-4}$ & 1.037 \\ \hline
$1.8 < Y < 2.6$  & $63.5^{+2.5}_{-1.9}{}^{+1.7}_{-2.0}$ & $63.7^{+2.5}_{-1.9}{}^{+1.7}_{-2.0}$ & 1.030 \\ \hline
\end{tabular}
\caption{The top-quark rapidity distribution at aN$^3$LO QCD and aN$^3$LO QCD$\times$EW, with $m_t=172.5$ GeV and MSHT20 aN$^3$LO pdf.
The central results are with $\mu=m_t$ and are shown together with scale and pdf uncertainties. The last column provides the aN$^3$LO QCD over NNLO QCD $K$-factor.}
\label{tabley2}
\end{center}
\end{table}

\begin{table}[htb]
\begin{center}
\begin{tabular}{|c|c|c|c|c|} \hline
\multicolumn{4}{|c|}{Top-quark $Y$ distribution at 13 TeV with CT18 NNLO pdf } \\ \hline
$d\sigma/dY$ in pb & aN$^3$LO QCD & aN$^3$LO QCD$\times$EW & aN$^3$LO QCD / \\ &  &  &  NNLO QCD \\ \hline
$0 < Y < 0.45$  & $249^{+6}_{-4}{}^{+5}_{-6}$ & $248^{+6}_{-4}{}^{+5}_{-6}$ & 1.028 \\ \hline
$0.45 < Y < 0.9$  & $226^{+6}_{-4}{}^{+5}_{-5}$ & $225^{+6}_{-3}{}^{+5}_{-5}$ & 1.031 \\ \hline
$0.9 < Y < 1.35$  & $185^{+5}_{-3}{}^{+4}_{-4}$ & $185^{+5}_{-4}{}^{+3}_{-4}$ & 1.031 \\ \hline
$1.35 < Y < 1.8$  & $134^{+5}_{-2}{}^{+4}_{-2}$ & $134^{+5}_{-2}{}^{+4}_{-2}$ & 1.032 \\ \hline
$1.8 < Y < 2.6$  & $67.4^{+2.7}_{-1.9}{}^{+2.7}_{-1.8}$ & $67.6^{+2.7}_{-1.9}{}^{+2.7}_{-1.8}$ & 1.031 \\ \hline
\end{tabular}
\caption{The top-quark rapidity distribution at aN$^3$LO QCD and aN$^3$LO QCD$\times$EW, with $m_t=172.5$ GeV and CT18 NNLO pdf.
The central results are with $\mu=m_t$ and are shown together with scale and pdf uncertainties. The last column provides the aN$^3$LO QCD over NNLO QCD $K$-factor.}
\label{tabley3}
\end{center}
\end{table}

\begin{table}[htb]
\begin{center}
\begin{tabular}{|c|c|c|c|c|} \hline
\multicolumn{4}{|c|}{Top-quark $Y$ distribution at 13 TeV with NNPDF4.0 NNLO pdf } \\ \hline
$d\sigma/dY$ in pb & aN$^3$LO QCD & aN$^3$LO QCD$\times$EW & aN$^3$LO QCD / \\ &  &  &  NNLO  QCD \\ \hline
$0 < Y < 0.45$  & $245^{+6}_{-5}{}^{+2}_{-1}$ & $244^{+6}_{-5}{}^{+2}_{-1}$ & 1.028 \\ \hline
$0.45 < Y < 0.9$  & $221^{+5}_{-5}{}^{+1}_{-2}$ & $220^{+5}_{-5}{}^{+1}_{-1}$ & 1.027 \\ \hline
$0.9 < Y < 1.35$  & $181^{+5}_{-4}{}^{+1}_{-1}$ & $181^{+5}_{-4}{}^{+1}_{-1}$ & 1.038 \\ \hline
$1.35 < Y < 1.8$  & $131^{+3}_{-4}{}^{+1}_{-1}$ & $131^{+3}_{-4}{}^{+1}_{-1}$ & 1.038 \\ \hline
$1.8 < Y < 2.6$  & $63.9^{+2.3}_{-1.9}{}^{+0.5}_{-0.5}$ & $64.1^{+2.3}_{-1.9}{}^{+0.5}_{-0.5}$ & 1.034 \\ \hline
\end{tabular}
\caption{The top-quark rapidity distribution at aN$^3$LO QCD and aN$^3$LO QCD$\times$EW, with $m_t=172.5$ GeV and NNPDF4.0 NNLO pdf.
The central results are with $\mu=m_t$ and are shown together with scale and pdf uncertainties. The last column provides the aN$^3$LO QCD over NNLO QCD $K$-factor.}
\label{tabley4}
\end{center}
\end{table}

From the numerical values reported in the top-quark rapidity tables we note that, within the required precision, EW corrections modify in a visible way only the first and last bin: they induce a decrease in the cross section at low rapidity values and increase the cross section at large rapidity.

Also in this case, the CT18 NNLO pdf provides cross section values in every bin which are systematically bigger than the ones obtained from the other pdf sets, while the MSHT20 aN$^3$LO pdf provides the smallest ones. 
Nonetheless, the $K$-factors are practically the same for all four pdf sets used in this calculation. As in the case of the $p_T$ distributions, the rapidity distributions obtained with the CT18 NNLO, MSHT20 NNLO and MSHT20 aN$^3$LO pdf have similar and more conservative pdf uncertainties as compared to those from NNPDF4.0 pdf in the considered $Y$ bins.    

\subsection{Comparison with 13 TeV LHC top-rapidity data}
\label{Y-chi2}

In this section, we compare our theoretical predictions for the top-quark $Y$ distributions at NNLO QCD$\times$EW and aN$^3$LO QCD$\times$EW with 13 TeV dilepton data from CMS~\cite{CMS:2018adi} and lepton+jets data from ATLAS~\cite{ATLAS:2019hxz}. As in the previous section, we use $\mu=m_t$ for the central results. Also in this case, pdf uncertainties have been computed using \texttt{fastNLO} for the NNLO predictions~\cite{Czakon:2017dip} and the \texttt{fastNLO-toolkit-v2.5.0}~\cite{Kluge:2006xs,Wobisch:2011ij,Britzger:2012bs,Britzger:2015ksm}.

\begin{figure}[htbp]
\begin{center}
\includegraphics[width=140mm]{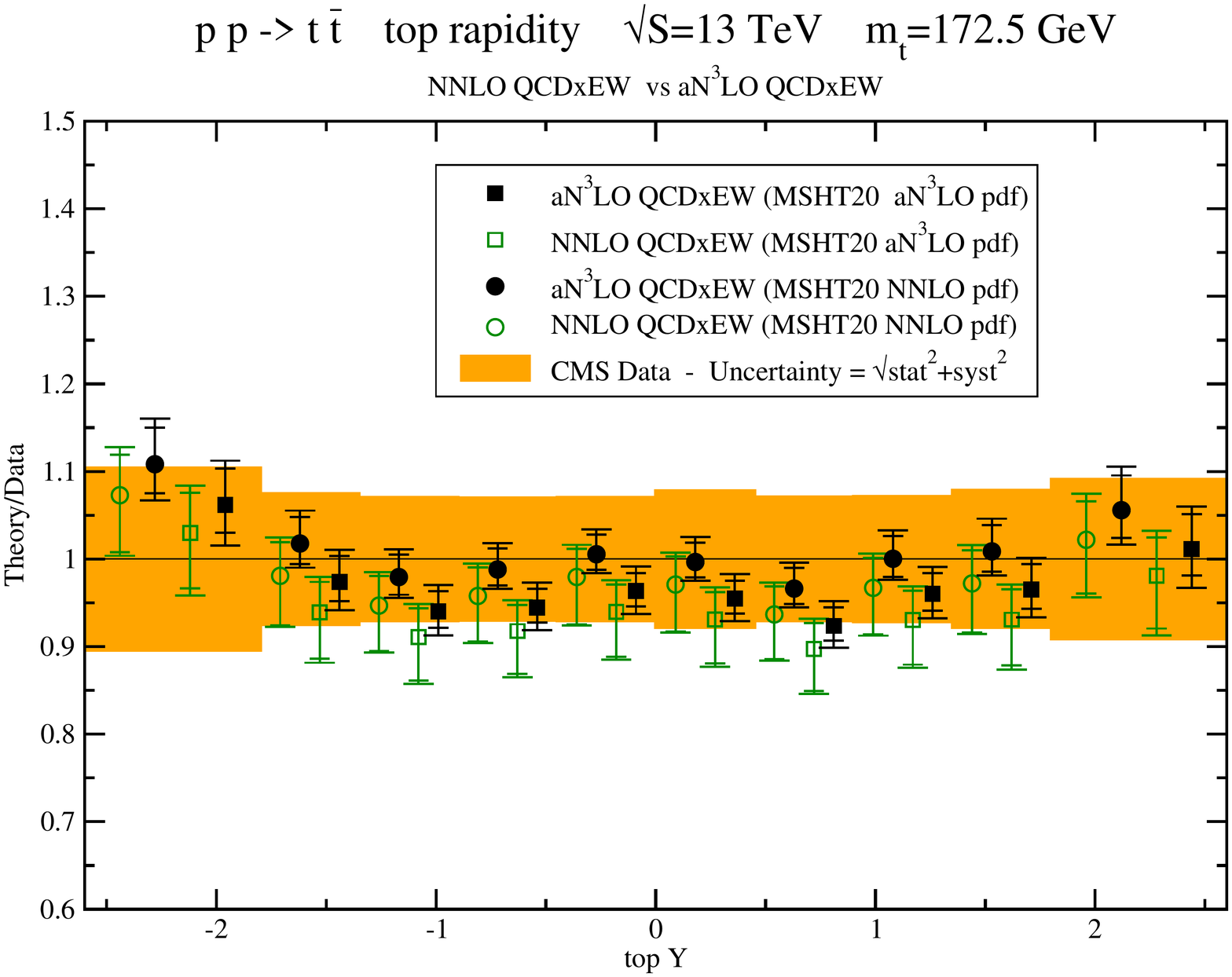}
\includegraphics[width=140mm]{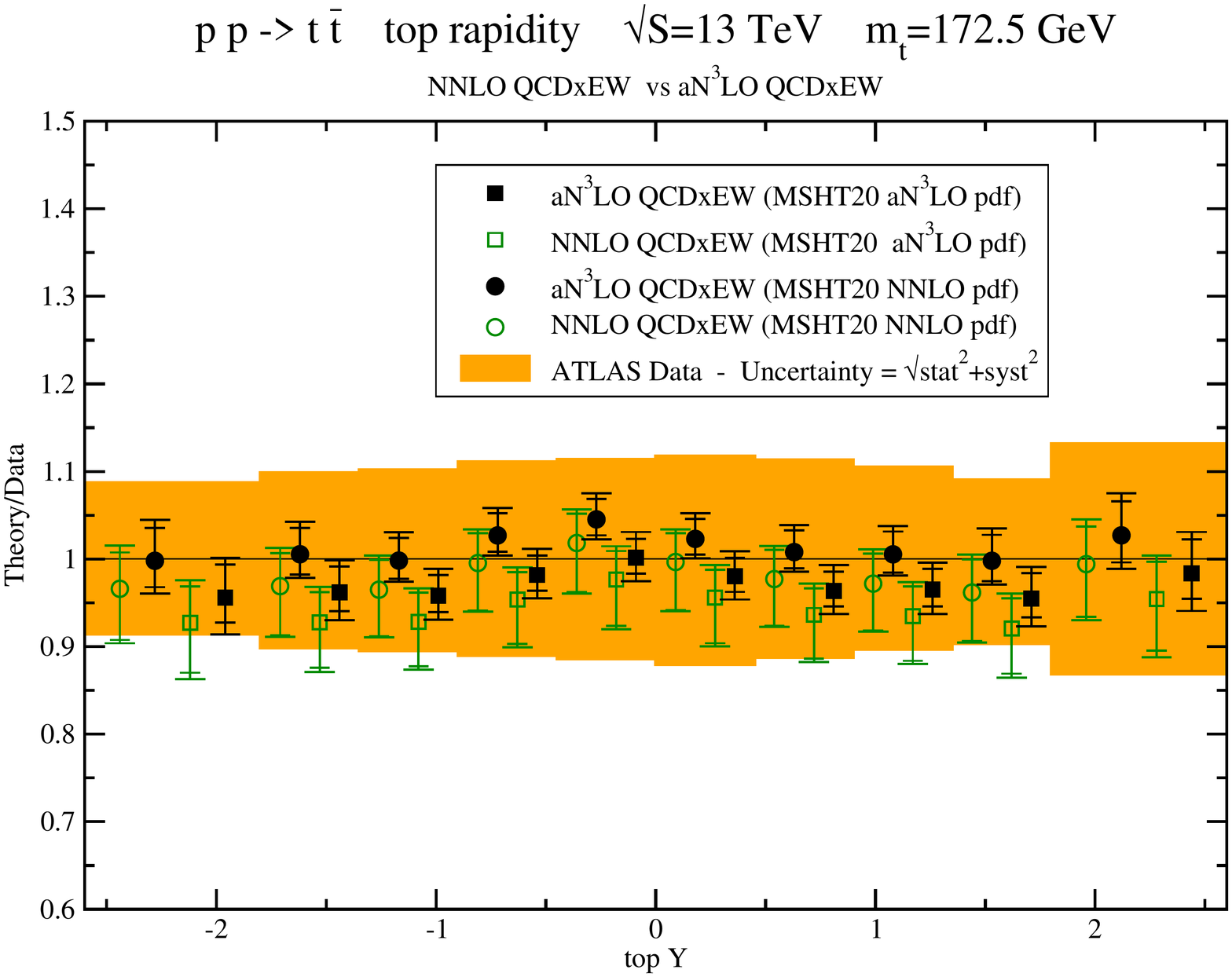}
\caption{Comparison of NNLO QCD$\times$EW and aN$^3$LO QCD$\times$EW theory predictions using MSHT20 NNLO and aN$^3$LO pdf with CMS (upper plot) and ATLAS (lower plot) top-quark rapidity data. The orange band represents the sum of statistical and systematic experimental uncertainties added in quadrature. Inner (outer) bars represent scale (scale plus pdf) theoretical uncertainties.}
\label{yt_msht_data_1_bar}
\end{center}
\end{figure}

\begin{figure}[htbp]
\begin{center}
\includegraphics[width=140mm]{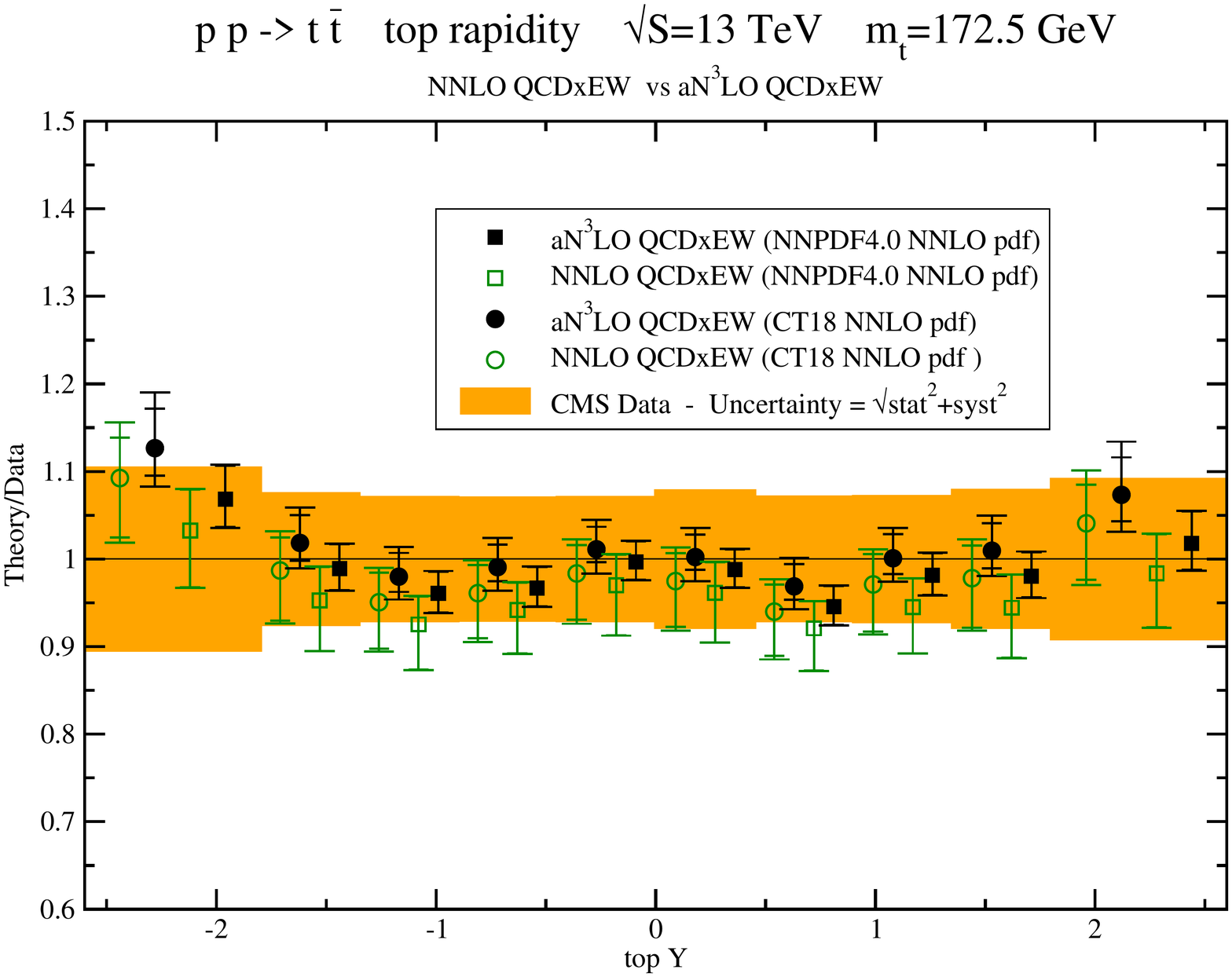}
\includegraphics[width=140mm]{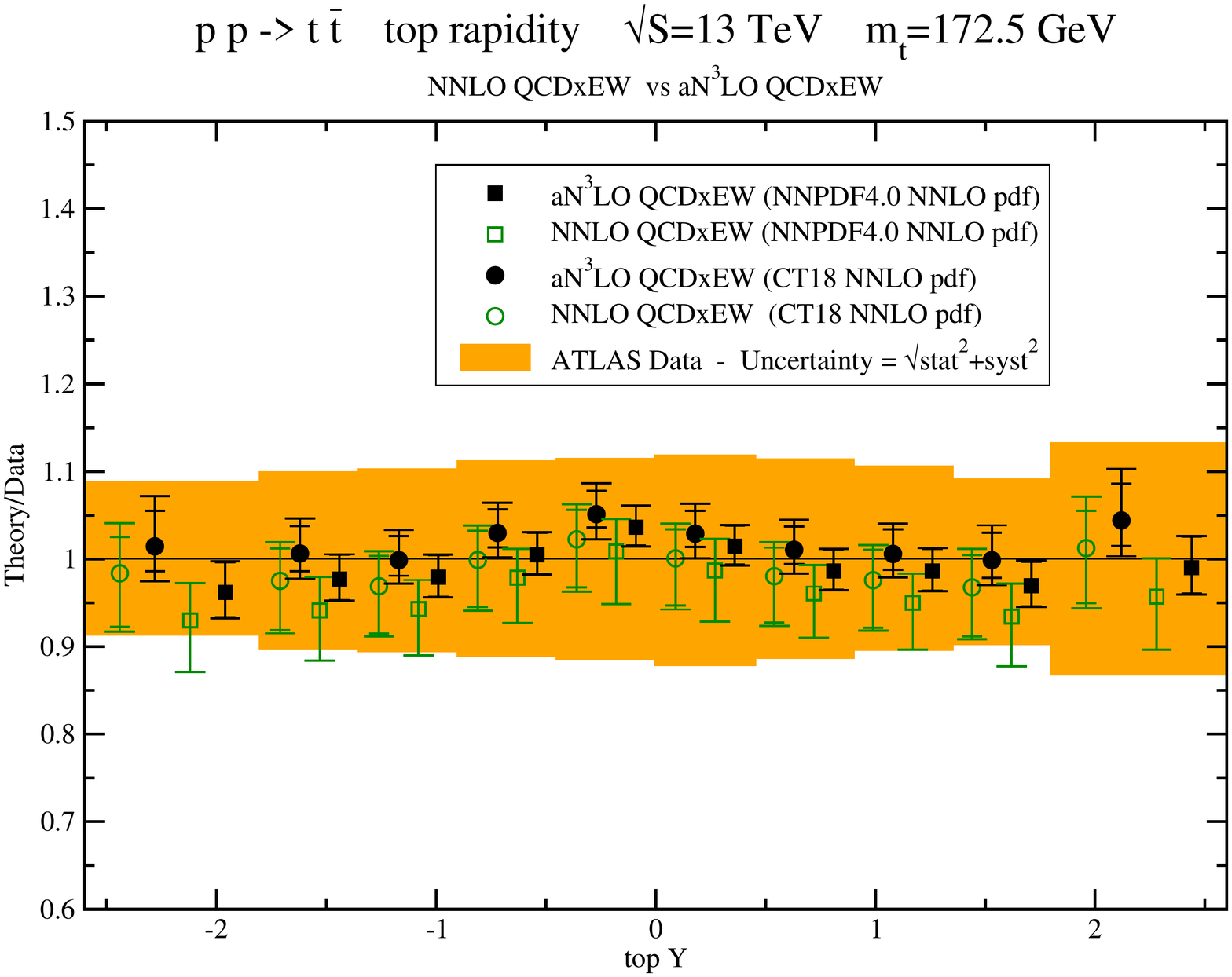}
\caption{Comparison of NNLO QCD$\times$EW and aN$^3$LO QCD$\times$EW theory predictions using CT18 NNLO and NNPDF4.0 NNLO pdf with CMS (upper plot) and ATLAS (lower plot) top-quark rapidity data. The orange band represents the sum of statistical and systematic experimental uncertainties added in quadrature. Inner (outer) bars represent scale (scale plus pdf) theoretical uncertainties.}
\label{yt_ct_data_1_bar}
\end{center}
\end{figure}
In Fig.~\ref{yt_msht_data_1_bar}, the top-quark $Y$ distributions at NNLO QCD$\times$EW and aN$^3$LO QCD$\times$EW obtained with 
MSHT20 NNLO and MSHT20 aN$^3$LO pdf are compared with ATLAS and CMS measurements. In analogy, in Fig.~\ref{yt_ct_data_1_bar}  we compare CMS and ATLAS data with the theoretical prediction for the differential distribution in top-quark rapidity at NNLO QCD$\times$EW and aN$^3$LO QCD$\times$EW obtained using CT18 NNLO and NNPDF4.0 NNLO pdf. In each plot we show the ratio of our theoretical predictions to the data together with scale and 68\% C.L. pdf uncertainties. Moreover, the orange band represents the experimental statistical and systematical uncertainties added in quadrature. From the distribution plots in Figs.~\ref{yt_msht_data_1_bar} and~\ref{yt_ct_data_1_bar} we note that the CMS and ATLAS measurements have differences, especially at large rapidity. 

Similarly to the $p_T$ case, we quantify the agreement of the theoretical predictions with the $Y$ distribution measured by CMS and ATLAS by computing $\chi^2/N_{pt}$ values for the theory predictions calculated at NNLO and aN$^3$LO, with and without EW corrections, using MSHT20 NNLO, MSHT20 aN$^3$LO, CT18 NNLO, and NNPDF4.0 NNLO pdf. In the CMS case we compute the $\chi^2$ using Eq. (\ref{chi2-CMS-formula}) and we summarize the values in Table~\ref{Y-chi2-CMS}.  In the ATLAS case we compute the $\chi^2$ using Eq. (\ref{chi2-ATLAS-formula}) and we summarize the values in Table~\ref{Y-chi2-ATL}. 

The theory predictions for the top-quark rapidity obtained by using NNPDF4.0 NNLO pdf give a better $\chi^2/N_{pt}$ in the case of the CMS measurements, while MSHT20 NNLO and MSHT20 aN$^3$LO pdf give a better $\chi^2/N_{pt}$ in the case of the ATLAS measurements. Once again, this reflects differences in these LHC data. 

We note that the inclusion of EW corrections in either NNLO or aN$^3$LO results increases the value of $\chi^2$ for the CMS data but decreases it for the ATLAS data. Also, the $\chi^2$ at aN$^3$LO is lower than at NNLO for both CMS and ATLAS data.

\begin{table}[!ht]
\begin{center}
\begin{tabular}{|c|c|c|c|c|}
\hline
  pdf & NNLO QCD & NNLO QCD   & aN$^3$LO QCD & aN$^3$LO QCD\\
        &                     &    $\times$EW &                           & $\times$EW\\
\hline
\hline
MSHT20 NNLO    & 0.71 & 0.76 & 0.66 & 0.70\\
MSHT20 aN$^3$LO  & 0.85 & 0.91 & 0.79 & 0.83\\
CT18 NNLO         & 0.86 & 0.92& 0.81& 0.88\\
NNPDF4.0 NNLO & 0.68 & 0.71& 0.56 & 0.61\\ 
\hline
\end{tabular}
\end{center}
\caption{Summary of the $\chi^2/N_{pt}$ for the top-quark rapidity distributions at CMS.}
\label{Y-chi2-CMS}
\end{table}

\begin{table}[!ht]
\begin{center}
\begin{tabular}{|c|c|c|c|c|}
\hline
  pdf & NNLO QCD & NNLO QCD   & aN$^3$LO QCD & aN$^3$LO QCD\\
        &                     &    $\times$EW &                           & $\times$EW\\
\hline
\hline
MSHT20 NNLO    & 0.70 & 0.66 & 0.49 & 0.44\\
MSHT20 aN$^3$LO  & 0.70 & 0.66 & 0.56& 0.46\\
CT18 NNLO         & 0.71 & 0.69& 0.71 & 0.70\\
NNPDF4.0 NNLO & 1.26 & 1.18& 0.90 & 0.84\\ 
\hline
\end{tabular}
\end{center}
\caption{Summary of the $\chi^2/N_{pt}$ for the top-quark rapidity distributions at ATLAS.}
\label{Y-chi2-ATL}
\end{table}

\subsection{Comparison with previous theoretical top-rapidity predictions}
\label{previous-top-Y}

Results at aN$^3$LO QCD for the top-quark rapidity distribution have been presented before with older pdf sets, beginning with Ref. \cite{NKan3lo2}. Several comparisons with older top-rapidity data from the LHC were presented in the review paper of Ref. \cite{NKtoprev}. Similarly to what we discussed in Sec.~\ref{previous-top-pT} for the $p_T$ distribution, those results were presented as functions of the top-quark rapidity (not as bins) and, thus, were not matched to the exact NNLO QCD result; instead, aNNLO and aN$^3$LO soft-gluon corrections were added to the NLO result. Since the binned aNNLO distributions are very close to the exact NNLO ones (at the few per mille level or better), the difference between the matched and unmatched aN$^3$LO distributions is negligible. We have checked that once again for the current pdf sets, but one can also see this in past results, e.g. in Ref. \cite{DPF2019} where the predictions are compared with the same CMS top-rapidity data as in this paper.

\section{Conclusions}
\label{Conclusion}
We have presented higher-order theoretical predictions for top-antitop pair production at the LHC. We have calculated the total cross section at aN$^3$LO in QCD, including EW corrections at NLO, taking into account scale and pdf uncertainties, for various LHC energies using pdf sets from the recent CT18 NNLO, MSHT20 NNLO, MHST20 aN$^3$LO, and NNPDF4.0 NNLO global analyses. 

We have also presented results for top-quark binned  differential distributions in transverse momentum and rapidity, including soft-gluon corrections through aN$^3$LO in QCD as well as electroweak corrections. These results have been obtained for collision energy of 13 TeV and are matched to exact NNLO QCD ones. Since the binned aNNLO distributions are very close to the exact NNLO ones, the difference between the matched and unmatched aN$^3$LO distributions is negligible. 

For both top-quark transverse-momentum and rapidity distributions we used the binning of Ref.~\cite{CMS:2018adi} and we compared with experimental results from CMS~\cite{CMS:2018adi} and ATLAS~\cite{ATLAS:2019hxz}. We quantified the quality of agreement of our theoretical predictions with the experimental data by computing $\chi^2/N_{pt}$. 
Overall, we found good agreement within the quoted uncertainties. We observed that the CMS and ATLAS top-quark $p_T$ and rapidity differential cross section measurements 
that we considered have differences which can potentially result in different kind of constraints (different pulls) on the gluon pdf at large $x$ in future global QCD analyses to determine pdf in the proton.   

In summary, the aN$^3$LO soft gluon corrections are important, they substantially increase the rates, and they decrease the scale uncertainties in the total cross sections as well as the $p_T$ and rapidity distributions. The electroweak corrections make significant contributions to the $p_T$ distribution, especially at large $p_T$, but their effect on the total cross section and the rapidity distribution is smaller. 
Our predictions contain the latest available theoretical input and are in good agreement with recent high-precision data from the ATLAS and CMS collaborations for total cross sections and differential distributions.

\section*{Acknowledgements}
This material is based upon work supported by the National Science Foundation under Grant No. PHY 2112025.

\end{document}